\documentclass[conference]{IEEEtran}
\IEEEoverridecommandlockouts
\usepackage{amsmath}
\usepackage{graphicx, subfigure}
\usepackage{amssymb}
\usepackage{cases}
\usepackage{cite}
\usepackage{url}
\usepackage[ruled,vlined,lined,ruled,linesnumbered]{algorithm2e}
\usepackage{relsize}
\usepackage[nodisplayskipstretch]{setspace}
\usepackage{longtable}
\usepackage{booktabs}
\usepackage{multirow}

\newtheorem{theorem}{Theorem}
\newtheorem{lemma}{Lemma}

\newtheorem{definition}{Definition}

\begin{document}
\title{Online Coordinated Charging Decision Algorithm for Electric Vehicles without Future Information \vspace{-0.3cm}}
\author{\IEEEauthorblockN{Wanrong Tang, Suzhi Bi and Ying Jun (Angela) Zhang,~\IEEEmembership{Senior Member,~IEEE}}
\IEEEauthorblockA{Department of Information Engineering, The Chinese University of Hong Kong\\
Shatin, New Territory, Hong Kong\\
Email:\{twr011,bsz009,yjzhang\}@ie.cuhk.edu.hk}
\thanks{This work was supported in part by the National Natural Science Foundation of China (Project number 61201261), the National Basic Research Program (973 program Program number 61101132) and the Competitive Earmarked Research Grant (Project Number $419509$) established under the University Grant Committee of Hong Kong.}

}
\maketitle

\begin{abstract}
The large-scale integration of plug-in electric vehicles (PEVs) to the power grid spurs the need for efficient charging coordination mechanisms.
It can be shown that the optimal charging schedule smooths out the energy consumption over time so as to minimize the total energy cost.
In practice, however, it is hard to smooth out the energy consumption perfectly, because the future PEV charging demand is unknown at the moment when the charging rate of an existing PEV needs to be determined.
In this paper, we propose an Online cooRdinated CHARging Decision (ORCHARD) algorithm,
which minimizes the energy cost without knowing the future information.
%we show that ORCHARD ensures finishing charging all PEVs and achieves the best known competitive ratio (2.39).
Through rigorous proof, we show that ORCHARD is strictly feasible in the sense that it guarantees to fulfill all charging demands before due time. Meanwhile, it achieves the best known competitive ratio of 2.39.
To further reduce the computational complexity of the algorithm, we propose a novel reduced-complexity algorithm to replace the standard convex optimization techniques used in ORCHARD.
Through extensive simulations,
we show that the average performance gap between ORCHARD and the offline optimal solution, which utilizes the complete future information, is as small as $14\%$.
By setting a proper speeding factor, the average performance gap can be further reduced to less than $6\%$.
\end{abstract}

\section{INTRODUCTION}
\subsection{Background and Contributions}\label{sec:background and contributino}
In recent years,
billions of dollars have been pledged to fund the development of electric vehicles and their components\cite{sovacool2009beyond}.
%At the mean time, the massive load caused by the integration of plug-in Electric Vehicles (PEVs) into the power grid has raised concerns about the potential cost to maintain the stability of power system \cite{fox2013getting}.
%It is shown in \cite{lopes2011integration} that both voltage instability and transmission congestion can arise from uncontrolled PEV charging.
At the mean time, the massive load caused by the integration of Plug-in Electric Vehicles (PEVs) into the power grid has raised concerns about the voltage instability and transmission congestion\cite{lopes2011integration}. Uncontrolled PEV charging will lead to potential cost at both generation and transmission sides.
To mitigate the negative effects and enjoy the benefit of PEVs' integration, it is critical to develop efficient charging control algorithms \cite{sortomme2011coordinated}.
Most of the existing PEV charging algorithms are ``offline'' in the sense that they rely on the non-causal information of future PEV charging profiles when deciding the charging schedules.
That is, the arrival time and charging demand of a PEV are assumed to be known to the charging station prior to the arrival of the PEV.
For instance, \cite{ma2010decentralized} requires all PEVs to negotiate with the charging station about their charging schedules one day ahead.
However, this assumption does not hold in practice.
A PEV's charging profile is revealed only after it arrives at the charging station or connects to the charging pole.

Consider the most general case where neither the PEV arrival instants, the charging demands, nor their distributions are known \emph{a priori}.
We are interested in developing an online charging algorithm that schedules PEV charging based only on the information of the PEVs that have already arrived at the charging station.
Specifically, online algorithm refers to a technique that processes inputs in a sequential manner with no requirement for future data \cite{allan98}.
There are some recent studies on online algorithms for PEV charging \cite{gerding2011online,clement2010impact,masoum2012distribution,he2012optimal,gan2012optimal}.
For instance, \cite{gerding2011online} designed an online auction protocol for PEVs to bid for charging opportunities.
However, it assumes that different types of PEVs have the same fixed charging rate, which is not true in practice.
\cite{clement2010impact,masoum2012distribution} proposed real-time charging strategies by assuming that all PEVs own the same plug-in time periods. In practice, however, the plug-in time varies from different PEV owners.
The algorithm proposed in \cite{he2012optimal} is based on an overly simplifying assumption that there will be no future PEV arrivals at the moments when charging schedules are decided.
The resulting charging schedule is suboptimal, as it underestimates the overall charging load.
%In this paper, we propose a simple yet effective speeding algorithm to improve upon the algorithm in \cite{he2012optimal}.
%On average $10\%$ performance gain is observed in simulations.
To date, most of the existing works, including \cite{gerding2011online,clement2010impact,masoum2012distribution,he2012optimal,gan2012optimal},
fail to provide any performance analysis of their online algorithms.
%\cite{chen2012iems} analyzed the performance but did not guarantee that all the PEVs can finish charging before their departures.
The few work that analyzed the performance, e.g., \cite{chen2012iems}, does not guarantee the satisfaction of PEVs' charging demands before their departures.

In this paper, we propose an efficient Online cooRdinated CHARging Decision (ORCHARD) algorithm that aims to minimize the total charging cost by mimicking the offline optimal charging decision. ORCHARD is strictly feasible in the sense that it guarantees to satisfy all PEVs' charging demands before their due time, as long as the charging demands are feasible.
In contrast to the algorithms proposed in \cite{ma2010decentralized,clement2010impact,masoum2012distribution}, ORCHARD allows heterogeneity among PEVs. That is, PEVs can have arbitrary arrival (or plug-in time) and departure times, charging demands and maximum charging rates. We show that ORCHARD is strictly feasible in the sense
that it guarantees to fulfill all charging demands before the due time.
More importantly, we rigorously analyze the performance of ORCHARD in terms of competitive ratio.
Our analysis shows that ORCHARD achieves a competitive ratio of $2.39$, which is the best known ratio so far.
%Indeed, the proposed algorithm not only solves the charging scheduling problem in PEV systems, but also generalizes the investigation of speed scaling problem, which has been a focus of study in the online algorithm literature \cite{bansal2009improved}.
To further reduce the computational complexity, we propose a low-complexity optimization routine to replace the standard convex optimization algorithms used in ORCHARD.
Extensive simulations show that the average performance gap between ORCHARD and the offline optimal solution is as small as $14\%$. The gap can be reduced to $6\%$, if the speeding factor used in the algorithm is carefully chosen according to the charging demand pattern.

\subsection{Related Work}\label{sec:related work}

The charging scheduling for PEV is similar to, but not the same as, the speed scaling problem in the CS literature. Speed scaling is a power management technique that involves dynamically changing the speed of a processor \cite{yao1995a,bansal2007speed,bansal2009improved,bansal2009speed,lam2009speed}. Specifically, the processor must schedule in real-time a number of tasks and allocate a processing rate to each of them, given that all tasks can be completed before their predetermined deadlines. The processor tries to minimize the total energy cost, where the energy cost at each time $t$ is a positive power function of the total processing rate $s(t)$ at that time, i.e. $s^\alpha(t)$. The key difference from a PEV charging problem is that speed scaling does not place a constraint on the maximum processing rate of each individual job as the PEV charging problem, where each PEV has a maximum charging rate. Another difference is that the cost function of PEV charging problem is a general polynomial instead of a positive power function. In other words, the speed scaling problem is a special case of the PEV charging problem in both objective and constraints.

The first offline optimal algorithm to solve the speed scaling problem was proposed by Yao, Demers and Shenker (YDS) \cite{yao1995a}. Later, \cite{yao1995a} proposed two online algorithms, i.e. Average Rate (AVR) and Optimal Available (OA). Conceptually, AVR processes a task at a rate equals to its average work load within its specified starting time and deadline independent of other tasks. The algorithm is proved to be $2^{\alpha-1}\alpha^\alpha$-competitive in \cite{yao1995a}. OA uses YDS to calculate the current optimal processing rate by assuming no more tasks will be released in the future, and its competitive ratio was proved to be $\alpha^\alpha$ in \cite{bansal2007speed}. Apparently, the OA solution is suboptimal, as it underestimates the future workload. To address the problem, \cite{bansal2009improved} proposed a qOA algorithm that scales up the processing rate of OA by a factor $q>1$. It also showed that qOA works better than OA and AVG in terms of competitive ratio. There are many follow-up works on extended topics, such as managing both temperature and power \cite{bansal2007speed}, minimizing the total flow plus energy \cite{bansal2009speed} \cite{lam2009speed}. The existing online algorithms cannot be directly applied to solve our problem, mainly because they do not consider the limits on the maximum processing speeds of the tasks. The problem investigated in this paper, on the other hand, can be viewed as a generalized speed scheduling problem. That is, the proposed algorithm and its analysis can be directly extended to solve the generalized speed scaling problem where individual processing rate bound applies.

The rest of the paper is organized as follows.
We construct the offline model in Section \ref{sec:model}.
In Section \ref{sec:online algorithm}, performance metric and online algorithm $OA$ are introduced, and the online algorithm ORCHARD is proposed and analyzed. A method used to reduce the complexity of ORCHARD is put forward in Section \ref{sec:offline algorithm}.
Simulation results are presented in Section \ref{sec:sim}. Finally, the paper is concluded in Section \ref{sec:conclusions}.

\section{Offline Optimal PEV Charging Scheduling}\label{sec:model}

\subsection{Problem Formulation}\label{sec:formulations}

We consider the PEV charging scheduling problem, in which
%$N(T)$ PEVs arrive at random instants with random charging demands that must be fulfilled before their departure time.
%System time starts from the arrival of the first PEV and ends with the last departure.
%We denote by $T$ the system time.
%The PEVs are indexed from $1$ to $N(T)$ according to their arrival order.
PEVs arrive at the charging station at random instants with random charging demands that must be fulfilled before their departure time. Suppose that $N$ PEVs arrive during a time period $T$, indexed from $1$ to $N$ according to their arrival order.
Notice that for a given time period $T$, $N$ itself is a random variable due to the random arrival of PEVs. Let $D_i, t_i^{(s)}, t_i^{(e)}$ denote the charging demand, arrival time, and departure time of PEV $i$, respectively.
Due to the battery constraint,
PEV $i$ can only be charged at a rate $x_{it}\in \left[0, U_i\right]$,
where $U_i$ is the maximum charging rate.
For the formulation to be meaningful, we assume that all the charging demands are feasible. That is, $D_i\leq \min \{ U_i (t_i^{(e)} - t_i^{(s)}), \zeta_i\}$ holds for all $i$, where $\zeta_i$ is the battery capacity of PEV $i$.

We assume that the total charging rate of all PEVs can always be met by the charging station at any time $t$.
%, i.e., $\sum_{i\in \mathcal{I}(t)}U_i$.
Let $\mathcal{I}_t$ be the set of PEVs parking in the station at time $t$.
The charging station has the control of the charging rate $x_{it}$ for each PEV $i$. We define $s_t$ as the total charging rate of time $t$, i.e., $s_t = \sum_{i \in \mathcal{I}_t}x_{it}$.
The power source of a charging station range from fuel, boiler, turbine to generator.
From previous studies \cite{kothari03}, the generation cost at time $t$ is in general a quadratic function of the total charging rate at that time, i.e.,
\begin{equation}
as_t + bs_t^2
,
\end{equation}
%\begin{equation}
%a(\sum_{i \in \mathcal{I}_t}x_{it}) + b(\sum_{i \in \mathcal{I}_t}x_{it})^2
%,
%\end{equation}
%\begin{align}\label{costfunc}
%\small
%a(\sum_{i \in \mathcal{I}_t}x_{it}) + b(\sum_{i \in \mathcal{I}_t}x_{it})^2 ~\text{(C\$/hour)},
%\end{align}
where $a$ and $b$ are constants.
The optimal charging scheduling problem that minimizes the total cost is then formulated as follows:
\begin{subequations}
\label{model1}
%\small
\begin{align}
& \min_{x_{it}} & & \quad \int_0^{T} \left(a \sum_{i \in \mathcal{I}_t}x_{it} + b (\sum_{i \in \mathcal{I}_t}x_{it})^2 \right) \mathrm d t \label{model2}
\\
&\text{s. t. }  & & \quad \int_{t_i^{(s)}}^{ t_i^{(e)}} x_{it} \mathrm dt \geq D_i, i = 1,2,\ldots,N, \label{model3}
\\
&  & & \quad 0 \leq x_{it} \leq U_i, i = 1,2,\ldots,N, t\in \left[t_i^{(s)}, t_i^{(e)} \right], \label{model4}
%\\
%&  & & \quad x_{it} \leq U_i, i = 1,2,\ldots,N, t\in \left[t_i^{(s)}, t_i^{(e)} \right]. \label{model5}
\end{align}
\end{subequations}
Problem (\ref{model1}) is a convex optimization problem. In the ideal case when all PEVs' charging profiles, including $t_i^{(s)}$, $t_i^{(e)}$, $U_i$ and $D_i$, are known to the charging station noncausally at time $0$, one can obtain the optimal $x_{it}$ for all $i$ and $t$  by solving (\ref{model1}) before the start of system time. We refer to the optimal solution obtained with noncausal information as offline optimal solution. In practice, however, a PEV's charging profile is revealed only after it arrives at the station. In the section \ref{sec:online algorithm}, we will investigate an online PEV charging problem that determines the charging rate at each time $t$ based only on the current and past information.
\begin{figure}
%\vspace{-0.3cm}
        \centering
        \includegraphics[width=8cm,height=3.5cm]{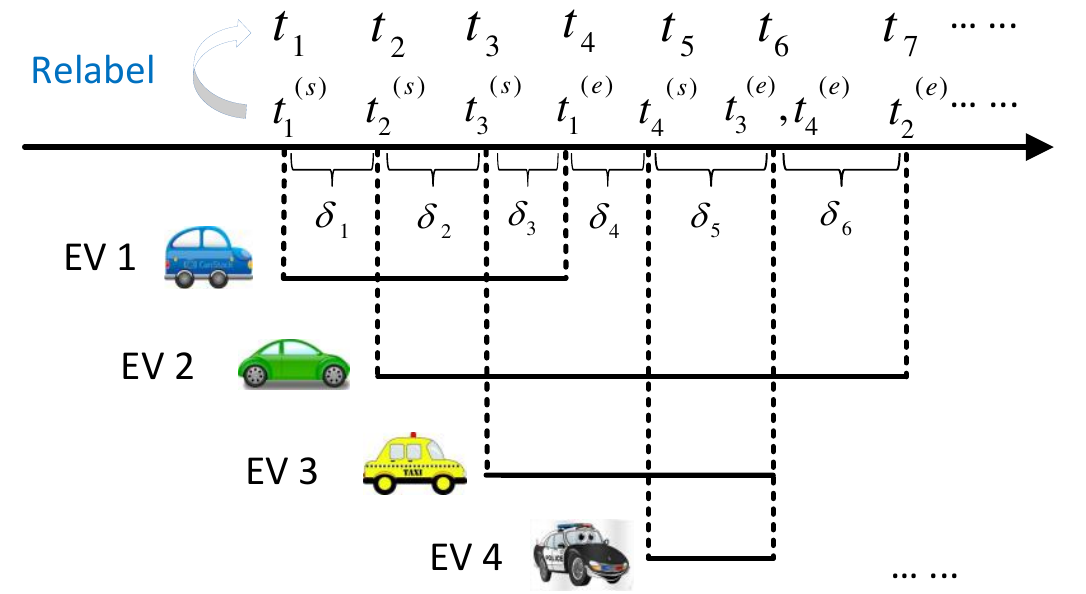}
        \caption{Illustration of one offline case}
        \label{fig:onlinemodel1}
%        \vspace{-0.3cm}
\end{figure}

\subsection{Model Transformation}
A close look at (\ref{model1}) suggests that there are infinite variables $x_{it}$, because the time $t$ is continuous. In this subsection, we show that the problem can be equivalently transformed to one with a finite number of variables.

%The conversion would be helpful to exploit the solution structure, which eventually leads to an efficient online algorithm without losing any precision.

As illustrated in Fig.~\ref{fig:onlinemodel1},
%the time instances $t_1,t_2,...,t_{K+1}$ are the indexed arrival and departure time of all PEVs. The $K$ time intervals specified by the $K+1$ time instances are indexed by $\delta_1, \delta_2, ..., \delta_K$.
we relabel the time instants $t_i^{(s)}$ and $t_i^{(e)}$ as $t_1,t_2,\cdots$ in sequential order. Notice that the time instants are the times when either an arrival or a departure event occurs.
It is possible that more than one PEV arrives or departs at the same time. For instance, in Fig.~\ref{fig:onlinemodel1}, both PEV3 and PEV4 leave at $t_6$. Define a time interval as the time period between two adjacent time instants. Notice that the set of cars parked in the station does not change in the middle of a time interval. Let $\mathcal{K}$ denote the set of indices of the time intervals, and
%$\delta_k (k \in \mathcal{K})$ denote the $k^{th}$ interval, and $|\delta_k|$ be the length of $\delta_k$.
$\delta_k (k \in \mathcal{K})$ denote the length of $k^{th}$ interval.
We show in Lemma \ref{lemma:modeltransform} that the total charging rate is a constant during each time interval.

\begin{lemma}\label{lemma:modeltransform}
%Let $x_{it}^*$ denote the optimal charging schedule that minimizes the total charging cost in (\ref{model1}).
Let $x_{it}^*$ denote an optimal solution to (\ref{model1}) and $s_t^* = \sum_{i \in \mathcal{I}_t} x^*_{it}$.
Then,
%it holds that $\sum_{i \in \mathcal{I}_t} x^*_{it}= s_{k}$, $\forall t\in\left[t_k,t_{k+1}\right)$, $k=1,...,K$, where $s_k$ is a constant number with $0 \leq s_{k} \leq \sum_{i \in \mathcal{I}(k)}U_i$.
the optimal total charging rate $s_t^*$ remains constant in each time interval. Moreover, there exists an optimal solution where $x^*_{it}$ remains constant during each time interval.
\end{lemma}

%The lemma can be easily proved by Jensen's inequality.
%We omit the proof here. Interested readers are referred to \cite{tang13report} for the detailed proof.
\emph{Proof:} The proof is given by contradiction. The optimal total charging rate at time $t\in\left[t_k,t_{k+1}\right)$ is denoted by $\widetilde{s}_{k}(t)=\sum_{i \in I(t_k)} x^*_{it}$, $k=1,...,K$. Let $s_k =  \int_{t_k}^{t_{k+1}} \widetilde{s}_{k} (t) \mathrm d t / (t_{k+1} - t_k )$
%\begin{equation}\label{theorem1proof1}
%\small
%\s_k = \frac{ \int_{t_k}^{t_{k+1}} \widetilde{s}_{k} (t) dt} { t_{k+1} - t_k }
%\end{equation}
be the average charging rate in $\delta_k$. Note that $s_k$ is always achievable by setting the charging rate of each EV $i$ as $\int_{t_k}^{t_{k+1}} x^*_{it} dt / \left(t_{k+1}-t_k\right)$. By Jensen's inequality, we have
\begin{equation}
%\small
\begin{aligned}
&\int_{t_k}^{t_{k+1}} \frac{a\hat{s}_{k}(t) + b \left(\hat{s}_{k}(t)\right)^2}{t_{k+1}-t_k}  dt \\
\geq& a \frac{ \int_{t_k}^{t_{k+1}} \hat{s}_{k} (t) dt} { t_{k+1} - t_k } + b \left[\frac{ \int_{t_k}^{t_{k+1}} \hat{s}_{k} (t) dt} { t_{k+1} - t_k } \right]^2 = a s_k + bs_k^2.\\
\end{aligned}
\end{equation}
Equivalently, we have
\begin{equation}
\label{1}
\begin{aligned}
%\small
\int_{t_k}^{t_{k+1}} \left [ a\hat{s}_{k}(t) + b\left(\hat{s}_{k}(t)\right)^2 \right ]dt \geq \left(t_{k+1} - t_k\right) (as_k + bs_k^2).
\end{aligned}
\end{equation}
From (\ref{1}), the uniform total charging rate $s_k$ incurs no higher cost than that of $x^*_{it}$, which contradicts with the assumption that $x^*_{it}$ is the optimal charging schedule. Therefore, the optimal charging schedule must produce constant total charging rate in each interval $\delta_k$, which completes the proof. \hfill $\blacksquare$

%Without loss of generality, we assume that the charging rate of each PEV $i$ in $\delta_k$ is a constant as well, denoted by $x_{ik}$.
Due to Lemma \ref{lemma:modeltransform}, we denote the constant charging rate of PEV $i$ in the $k^{th}$ interval by $x_{ik}$.
%Notice that this assumption does not affect the objective of (\ref{model1}), as long as the total charging rate remains the same.
Likewise, denote $\mathcal{J}(i)$ as the set of indices of the time intervals during which PEV $i$ parks in the station, $\mathcal{I}(k)$ as the set of PEVs that park in the $k^{th}$ interval.
%and $\delta_k$ as the duration of the $k^{th}$ interval.
Thanks to Lemma \ref{lemma:modeltransform}, we can equivalently transform problem (\ref{model1}) to the following form that has finitely many variables:
\begin{subequations}
\label{model6}
%\small
\begin{align}
& \min_{x_{ik}} & & \quad \sum_{k \in \mathcal{K}} \left (a \sum_{i \in \mathcal{I}(k)} x_{ik} + b(\sum_{i \in \mathcal{I}(k)} x_{ik})^2 \right) \delta_k \label{model7}
\\
& \text{s.t.} & & \quad \sum_{k \in \mathcal{J}(i)} x_{ik} \delta_k \geq D_i, i = 1,2,\ldots,N, \label{model8}
\\
&  & & \quad 0 \leq x_{ik} \leq U_i, i = 1,2,\ldots,N, k \in \mathcal{J}(i), \label{model9}
%\\
%&  & & \quad x_{ik} \leq U_i, i = 1,2,\ldots,N, k \in \mathcal{J}(i). \label{model10}
\end{align}
\end{subequations}

\section{Online Algorithm}\label{sec:online algorithm}

In this section, we formulate the online PEV charging problem and present an efficient online algorithm ORCHARD. We show that ORCHARD achieves a competitive ratio that is the best known so far. Moreover, the algorithm is strictly feasible in the sense that it always ensures to satisfy all PEV charging demands.

\subsection{Online PEV Charging and Performance Metric}\label{sec:online charging}

The online PEV charging problem assumes that, at any time instant $t$, the scheduler only knows the charging profiles of the PEVs that have arrived upon or before $t$. Based on the causal information, the scheduler makes an online decision of the charging rates $x_{it}$ at each time $t$. Once the decision is made, it cannot be changed afterwards. Without knowing the entire information, an online algorithm is forced to make decisions that may later turn out to be suboptimal. Thus, we have $\Psi_{ON}\geq \Psi^*$, where $\Psi_{ON}$ denotes the total cost induced by an online algorithm and $\Psi^*$ denotes the optimal cost obtained by the offline optimization.

A typical metric to evaluate the performance of an online algorithm is through competitive analysis, which compares the relative performance of an online and offline algorithm under the same sequence of inputs (e.g., the PEV charging profiles in our problem)\cite{allan98}. In particular, the competitive ratio of an online algorithm is the maximum ratio between its performance and that of the offline optimal algorithm over all possible input sequences. The formal definition is given in the following Definition \ref{def:competitive ratio} \cite{allan98}.
\begin{definition}\label{def:competitive ratio}
An online algorithm is $c-$competitive if there exists a constant $\theta$ such that
\begin{eqnarray}\label{defcom}
 \Psi_{ON} \leq c \cdot \Psi^* + \theta
\end{eqnarray}
holds for any input.
\end{definition}

The competitive ratio is always larger than $1$. Notice that the competitive ratio measures the performance ratio in the worst case. Very often, the average performance ratio is much smaller than $c$. This will be shown in the simulation section in which ORCHARD with competitive ratio equals 2.39 achieves an average performance ratio less than $1.06$.

\subsection{Online Optimal Available (OA) Algorithm}\label{sec:online oa}
In this subsection, we describe an intuitive online scheme called OA, which, although suboptimal, will be helpful in understanding our proposed ORCHARD algorithm.

%At a tagged time instant $\hat{t}_m$, the scheduler knows the set of cars parking at the station, i.e. $\mathcal{I}(\hat{t}_m)$, the residual demand to be satisfied for PEV $i$, denoted by $\bar{D}_{ij}$, and the deadline $t_i^{(e)}$ of PEV $i$ $\forall i \in \mathcal{I} (\hat{t}_m)$. Let $T_m$ denote the time instant the the latest departure time over all current PEVs, i.e., $T_m = \max \{ t_i^{(e)}: i \in \mathcal{I} (\hat{t}_m) \}$. Based on information at time instant $\hat{t}_m$, we formulate the following problem:

The OA algorithm works as follows. At a time instant $t_j$ when a PEV arrives, the scheduler calculates the optimal charging schedule assuming that no more PEVs will arrive in the future. More specifically, the scheduler solves the following problem (\ref{model11}) at time instant $t_j$, where $\mathcal{\bar{I}}(t, t_j)$ denotes the set of PEVs who have arrived by time $t_j$ and will be present at the future time $t \in (t_j, \bar{T}(t_j)]$,
\begin{equation}
\bar{T}(t_j) = \max \{ t_i^{(e)}: i \in \mathcal{I}_{t_j} \}
\end{equation}
denotes the latest departure time of all PEVs that have already arrived by $t_j$(recall that $\mathcal{I}_{t_j}$ is the set of PEVs parking in the station at
time $t_j$), $\bar{D}_i(t_j)$ denotes the residual demand to be satisfied for PEV $i$ at time $t_j$.
\begin{subequations}
\label{model11}
%\small
\begin{align}
& \min_{x_{it}} & & \quad \int_{t_j}^{\bar{T}(t_j)} \left ( a\sum_{i \in \mathcal{\bar{I}}(t, t_j)} x_{it} + b(\sum_{i \in \mathcal{\bar{I}}(t, t_j)} x_{it})^2 \right)\mathrm d t \label{model12}
\\
&\text{s. t. }  & & \quad \int_{t \in [t_j, t_i^{(e)}]} x_{it} \mathrm d t \geq \bar{D}_i(t_j), i \in \mathcal{I}_{t_j}, \label{model13}
\\
&  & & \quad 0 \leq x_{it} \leq U_i, i \in \mathcal{I}_{t_j}, t\in \left[t_j, t_i^{(e)} \right], \label{model14}
%\\
%&  & & \quad x_{it} \leq U_i, i = 1,2,\ldots,N, t\in \left[t_k, t_i^{(e)} \right]. \label{model15}
\end{align}
\end{subequations}

\begin{figure}
        \centering
        \includegraphics[width=8cm,height=3.5cm]{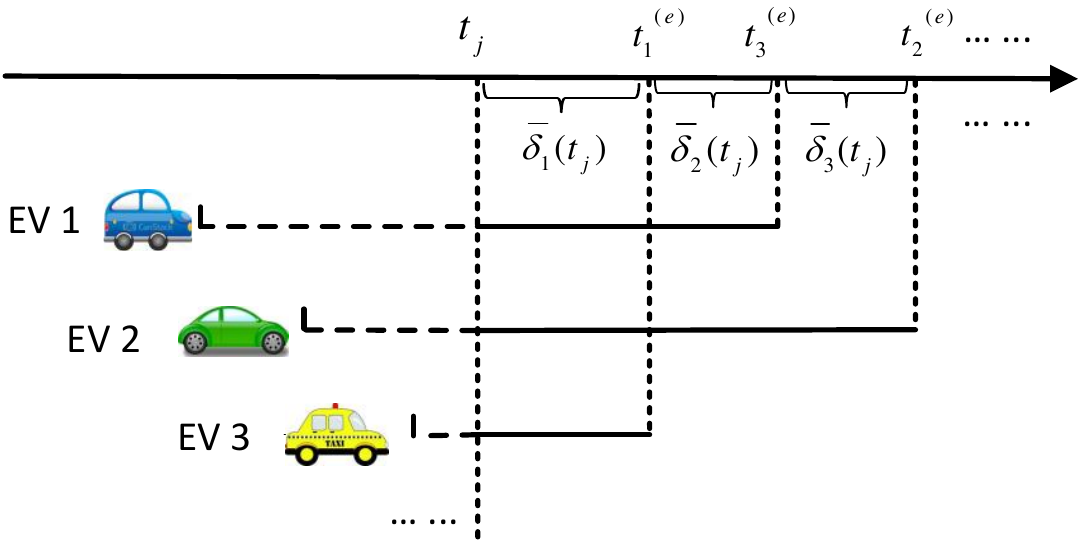}
        \caption{Illustration of one online case}
        %\vspace{-0.75cm}
        \label{fig:onlinemodel2}
\end{figure}

Notice that Problem (\ref{model11}) does not schedule the charging rates before time $t_j$. This is because the changing schedule that has been executed in the past cannot be changed afterwards. Having obtained the solution to (\ref{model11}), the scheduler charges the PEVs according to the solution until a new PEV arrives. Then, Problem (\ref{model11}) is re-solved with the updated set of charging profiles. Similar to the discussion in subsection II-B, the time axis in Problem (\ref{model11}) can be divided into intervals defined by the arrival and departure instants of the existing PEVs. By keeping a charging rate in each interval constant, Problem (\ref{model11}) can be equivalently transformed to one with finitely many variables.
%The equivalent formulation is similar to (\ref{model6}), except that the time intervals are defined by existing PEVs and the charging demand $D_i$ is replaced by residue charging demand $\bar{D}_i(t_j)$.
%Intuitively, (\ref{model11}) schedules charging assuming no future arrivals. Similar to the definition of time instants in Fig.~$\ref{fig:onlinemodel1}$, we illustrate the real-time charging profiles in Fig.~$\ref{fig:onlinemodel2}$, where the dotted transverse lines means the past time that PEVs parked. The departure time of each PEV $i\in \mathcal{I}_{t_j}$ is marked in the sequence of its departures, denoted by $t_{j1},t_{j2},...$. The definition of time intervals are also the same as Fig.~$\ref{fig:onlinemodel1}$ that is the time period in between two adjacent time instants. We label the intervals $1, 2, ...$ according to their orders. At the point of time $t_j$, we have following definitions.
As OA assumes no future arrivals, the time intervals are defined by time $t_j$ and the departure times of the PEVs that are present at $t_j$, as shown in Fig.~$\ref{fig:onlinemodel2}$.
Denote $\mathcal{\bar{K}}(t_j)$ as the set of indices of the intervals seen at time $t_j$, $\bar{\delta}_k(t_j)$, where $k \in \mathcal{\bar{K}}(t_j)$ as the length of the $k^{th}$ interval, $\mathcal{\bar{I}}(k,t_j)$ as the set of PEVs who have arrived by time $t_j$ and will be in the station at interval $k, k \in \mathcal{\bar{K}}(t_j)$, and $\mathcal{\bar{J}}(i,t_j)$ as the set of indices of time intervals that PEV $i$ will park in the station. It directly follows from Lemma \ref{lemma:modeltransform} that there exists an optimal solution to (\ref{model11}) where the optimal charging rates are constants during each interval.
%the optimal charging schedule solution of (\ref{model11}) also produces constant sum charging rate in each interval $k, k \in \mathcal{\bar{K}}(t_j)$.
Denote $x_{ik}$ by the charging rate of PEV $i$ in interval $k, k \in \mathcal{\bar{K}}(t_j)$. Then, (\ref{model11}) is equivalent to the following discrete time optimization problem
\begin{subequations}
\label{model16}
%\small
\begin{align}
& \min_{x_{ik}} & & \quad \sum_{k \in \mathcal{\bar{K}}(t_j)} \left ( a\sum_{i \in \mathcal{\bar{I}}(k,t_j)} x_{ik} + b(\sum_{i \in\mathcal{\bar{I}}(k,t_j)} x_{ik})^2 \right) \bar{\delta}_k(t_j) \label{model17}
\\
& \text{s.t.} & & \quad \sum_{k \in \mathcal{\bar{J}}(i,t_j)} x_{ik}\bar{\delta}_k(t_j) \geq \bar{D}_i(t_j), i \in \mathcal{I}_{t_j}, \label{model18}
\\
&  & & \quad 0 \leq x_{ik} \leq U_i, i \in \mathcal{I}_{t_j}, k \in \mathcal{\bar{J}}(i,t_j), \label{model19}
%\\
%&  & & \quad x_{ik} \leq U_i, i = 1,2,\ldots,N, k = 1, 2,\ldots, \kappa_i^{(e)}. \label{model20}
\end{align}
\end{subequations}

%Online OA algorithm is that once the charging profile changes, that is, a new PEV comes, update the current information, and reformulate the problem (\ref{model16}) and solve it.
%Each new PEV's arrival will trigger the online OA algorithm to update the available charging profiles. In this case, the OA algorithm reformulates (\ref{model16}) and calculates a new set of charging rate.
\subsection{The ORCHARD Algorithm}\label{sec:online description}
The charging rate scheduled by OA tends to be smaller than the optimal offline solution due to the neglection of future demands.
In ORCHARD, we speed up the charging schedule obtained from (\ref{model16}) by a speed-up factor $q$ ($q\geq 1$). Roughly speaking, the total charging rate by ORCHARD is $q$ times that of OA. The value of $q$ determines the performance of ORCHARD, including both the competitive ratio and the average performace. We will discuss how to set a proper $q$ to obtain the minimum competitive ratio in Section \ref{sec:competitive} and to obtain the best average performance in Section \ref{subsec:discussion_q}.

We denote by $\bar{x}_{ik}(t_j)$ the charging rate of PEV $i \in \mathcal{I}_{t_j}$ computed by OA at time $t_j$,  $\hat{x}_{it}$ the charging rate of PEV $i$ at time $t$ computed by ORCHARD, and $\hat{s}_t$ the sum of $\hat{x}_{it}$ at time $t$.
%Whenever a new PEV arrives or an existing PEV finishes charging, we need to update the charging profiles of the current PEVs. A point to notice is that the speeding-up factor will result in PEVs finishing charging before their departure time while OA always finish charging at the departure time exactly. This somehow changes the demand constraints in (\ref{model18}), and thus requires to recalculate the charging scheduling problem. Denote $t_{j-1}$ as the previous time instant that recalculate the charging scheduling problem before time $t_j$. Note that $\hat{x}_{it}$ is a constant between $t_{j-1}$ and $t_j$, where $\hat{x}_{it} = \hat{x}_{it_{j-1}}, \forall t \in [t_{j-1}, t_j), j = 2, 3, \cdots$.
%In particular, the charging demand of each PEV $i$, in constraint (\ref{model18}), is updated by
With the speed-up factor $q$, the OA problem (Problem (\ref{model11}) or (\ref{model16})) needs to be re-solved not only when there is a new car arrival, but also when a PEV finishes charging. This is because the PEVs may finish charging before their departure time calculated by OA due to the speedup.
At each time instant $t_j$ when the OA problem has to be re-solved, the right hand side of (\ref{model18}) is updated as follows
\begin{equation} \label{model21}
\small
\begin{aligned}
&\bar{D}_i(t_j) \\
&= \begin{cases}
0, &   \text{if PEV $i$ finishes charging,} \\
D_i, & \text{if PEV $i$ arrives,} \\
\bar{D}_i(t_{j-1}) - \hat{x}_{it_{j-1}}(t_j - t_{j-1}), & \text{otherwise.}
\end{cases}
\end{aligned}
\end{equation}
Here, $\hat{x}_{it_{j-1}}$ denotes the constant charging rate between $t_{j-1}$ and $t_j$.
%Moreover, we also need to update the set of intervals that current PEVs park on, which corresponds to $\mathcal{\bar{J}}(i,t_j)$ in constraint (\ref{model19}), interval length $\bar{\delta}_k(t_j)$ and the set of PEVs that park on each updated interval, i.e., $\mathcal{\bar{I}}(k,t_j)$. Then, we can obtain a new charging schedule by solving an updated (\ref{model16}) in respond to the change of charging profiles. Notice that only the charging schedule $\bar{x}_{i1}(t_j)$ for the time interval $\bar{\delta}_1(t_j)$ will be used as the current charging rate for each PEV. This is because there must be at least one PEV finishing charging at (or before) $t_{j1}$, such that the demand constraint of problem (\ref{model16}) is changed.
Moreover, we also need to update $\mathcal{\bar{J}}(i,t_j)$, $\bar{\delta}_k(t_j)$ and $\mathcal{\bar{I}}(k,t_j)$ according to the PEVs that are present at $t_j$.
A pseudo code of ORCHARD is presented in Algorithm \ref{alg:orchard} and explained as follows:

\textit{Step 1: } Once a PEV arrives or finishes charging, at time $t_j$, we set $t_j$ as the current starting time and update $\mathcal{I}_{t_j}$ as well as other parameters based on current information of PEVs (line 3).

\textit{Step 2: } Calculate the charging solution $\bar{x}_{i1}(t_j)$ for PEV $i\in\mathcal{I}_{t_j}$ using the OA algorithm, i.e. solving (\ref{model16}) by given the updated parameters (line 4).

\textit{Step 3: } Determine the total charging rate, which is the minimum of $q$ times of the total charging rate computed by OA, i.e., $ \sum_{i \in  \mathcal{\bar{I}}_{t_j}} \bar{x}_{i1}(t_j)$, and the sum of maximum charging rate of current PEVs, i.e., $\sum_{i \in \mathcal{\bar{I}}_{t_j}} U_i $ (line 5).

\textit{Step 4: } Determine the charging solution of ORCHARD, by setting the charging rate of PEV $i$ as in line 6 in Algorithm~\ref{alg:orchard}, we can make sure: 1) for each PEV, the charging rate does not exceed its maximum charging rate, i.e., $x_{it} \leq u_i, i \in \mathcal{I}_{t_j}$; 2) the sum of the charging rates equals total charging rate given by Step 3, i.e., $\sum_{i \in \mathcal{I}_{t_j}} \hat{x}_{it} = \hat{s}_t$; 3) for each PEV, the charging rate is no smaller than the solution given by OA in Step 2, i.e., $\hat{x}_{it} \geq \bar{x}_{i1}(t_j), \forall i \in \mathcal{I}_{t_j}$ (line 6).
%There may be a variety of solutions which satisfy the three conditions. We choose an intuitive one and charge the current PEVs (line 6).

%\vspace{-0.1cm}
\begin{algorithm}\label{alg:orchard}
%\small
\SetAlgoLined
 \SetKwData{Left}{left}\SetKwData{This}{this}\SetKwData{Up}{up}
 \SetKwRepeat{doWhile}{do}{while}
 \SetKwFunction{Union}{Union}\SetKwFunction{FindCompress}{FindCompress}
 \SetKwInOut{Input}{input}\SetKwInOut{Output}{output}
 \Input{$U_i$, $t_i^{(e)}$, $D_i$ of all parking PEVs}
 \Output{$\hat{x}_{it}$}
 initialization $j=0$\;
\While{a PEV arrives or finishes charging}{
Let $j=j+1$, record current time $t_j$. Calculate $\bar{\delta}_k(t_j)$, $\mathcal{\bar{I}}(k,t_j)$, $k \in \mathcal{\bar{K}}(t_j)$, $\mathcal{\bar{J}}(i,t_j)$, $\bar{D}_i(t_j)$, $i \in \mathcal{I}_{t_j}$.\\
Solve problem (\ref{model16}) for the optimal solution $\bar{x}_{i1}(t_j) \forall i \in \mathcal{\bar{I}}_{t_j}$.\\
%Solve problem (5) for the optimal solution $x_{i1}^\emph{a} \forall i \in I(t_k)$.\\
Set $\hat{s}_t = \min \{ q \cdot \sum_{i \in  \mathcal{\bar{I}}_{t_j}} \bar{x}_{i1}(t_j), \sum_{i \in \mathcal{\bar{I}}_{t_j}} U_i \}$.\\
Set the charging rate of PEV $i$ at the time $t \geq t_j$ as $\hat{x}_{it} = \min \{ \bar{x}_{i1}(t_j) + \frac{U_i - \bar{x}_{i1}(t_j)}{\sum_{i \in \mathcal{\bar{I}}_{t_j}} (U_i - \bar{x}_{i1}(t_j))} \cdot \frac{q - 1}{q} \hat{s}_t, U_i \}$.\\
}
\caption{ORCHARD} \label{onlinealgorithm}
\end{algorithm}
%\vspace{-0.3cm}
%ORCHARD can always produce a feasible solution to (\ref{model6}). That is, all PEVs can finish charging before their departures under bounded charging rate constraints. It's easy to verify for any $t \in [t_j, t_{j+1}), j = 1, 2, \cdots,$
%\begin{equation}\label{onlinefeasible5}
%\small
%\bar{x}_{i1}(t_j) \leq \hat{x}_{it} \leq U_i,\ \ \sum_{i \in \mathcal{I}_t} \hat{x}_{it} = \hat{s}_t.
%\end{equation}
%Meanwhile, the charging schedule $\hat{x}_{it}$ can finish the charging of all PEVs before their departures. This is because it is no slower than the optimal charging schedule $\bar{x}_{i1}(t_j)$, which guarantees the feasibility of (\ref{model6}).
Note that OA always guarantees a feasible solution. It can be easily inferred that ORCHARD can also guarantee producing a feasible solution since its charging rate is always no smaller than that of the OA.
\subsection{Derivation of Competitive Ratio}\label{sec:competitive}
In this subsection, we show that ORCHARD is $2.39$-competitive.
Here, we use an amortized local competitiveness analysis and a potential function $\Phi(t)$ which is a function of time. In particular, $\Phi$ is chosen as
\begin{equation}
\Phi(0)=\Phi(T)=0.
\end{equation}
We always denote the current time as $\tau_0$. Let $\hat{s}$ and $s^*$ be the current total charging rate of ORCHARD and the optimal offline algorithm respectively. In order to establish that ORCHARD is $c$-competitive, it is sufficient to show that the following key equation
\begin{equation}\label{onlineratio1}
%\small
(a\hat{s} + b(\hat{s})^2) + {\mathrm d \Phi \over \mathrm d \tau_0} \leq c \cdot (as^* + b(s^*)^2),
\end{equation}
holds for all $\tau_0\leq T$.
This is because the integral over the entire time $T$ on both sides leads to
\begin{equation}\label{onlineratio2}
%\small
\int_0^{T} (a\hat{s} + b(\hat{s})^2) \mathrm d t \leq c \cdot \int_0^{T} (as^* + b(s^*)^2) \mathrm d t,
\end{equation}
where $\int_0^{T} (a\hat{s} + b(\hat{s})^2) \mathrm d t$ is the total cost of ORCHARD, $\int_0^{T} (as^* + b(s^*)^2) \mathrm d t$ is the cost of optimal offline algorithm. In this sense, (\ref{onlineratio2}) is consistent with the definition in (\ref{defcom}). Before providing the proof of competitiveness, we introduce the following notations. Let $\hat{w}(t', t'')$ and $w^*(t',t'')$ denote the total remaining demand of PEVs whose deadlines are between $[t',t'']$ for ORCHARD and the offline optimal algorithm respectively. We further denote
%$d(t',t'') = \max \{0,  \hat{w}(t', t'')- w^*(t', t'') \}$
\begin{equation}\label{onlineratio3}
%\small
\begin{aligned}
d(t', t'') = &\max \Big \{ 0, \min \{ \hat{w}(t', t''), \frac{1}{q} \sum_{i \in \mathcal{I}(t')}U_i (t'' - t') \}\\
& - \min \{ w^*(t', t''), \sum_{i \in \mathcal{I}(t')}U_i(t'' - t') \} \Big \}
\end{aligned}
\end{equation}
as the amount of additional demand left for ORCHARD with deadline in $(t',t'']$. Then, we define a sequence of time points $\tau_1, \tau_2,  \cdots$ as follows: let $\tau_1$ be the time such that $d(\tau_0,\tau_1)/(\tau_1 - \tau_0)$ is maximized. If there are several such points, we choose the furthest one. Given $\tau_k$, we let $\tau_{k+1} > \tau_k$ be the furthest point that maximizes $d(\tau_k, \tau_{k+1})/(\tau_{k+1} - \tau_k)$, i.e.,
\begin{equation}
\tau_{k+1} = \underset{\tau > \tau_k} {\mathrm{arg~max}} ~d(\tau_k, \tau)/(\tau - \tau_k).
\end{equation}
The ``load intensity gap'' within $(\tau_k, \tau_{k+1}]$ is denoted as
%$g_k = d(\tau_k, \tau_{k+1})/(\tau_{k+1} - \tau_k), k = 1,2,\cdots$.
\begin{eqnarray}\label{onlineratio4}
%\small
g_k = d(\tau_k, \tau_{k+1})/(\tau_{k+1} - \tau_k), k = 1,2,\cdots.
\end{eqnarray}
Evidently, $g_k$ is a non-negative monotonically decreasing sequence.

We are now ready to define the potential function $\Phi$ as
\begin{equation}\label{onlineratio6}
%\small
   \Phi= \beta_1 \cdot a \sum_{k = 0}^{\infty}((\tau_{k+1} - \tau_k) g_k) + \beta_2 \cdot b \sum_{k = 0}^{\infty}((\tau_{k+1} - \tau_k) g_k^2),
\end{equation}
where $\beta_1, \beta_2$ are constants which will be assigned values later. We notice that $\Phi(0)=\Phi(T)=0$ holds, since the load is clearly zero before any PEV arrives and after the last deadline.

Before we give the Theorem \ref{theormonline}, we provide the following Lemma which will be used in the theorem.
\begin{lemma}\label{lemmaonline}
%The inequality
\begin{equation}\label{onlineratio5}
qg_0 \leq \hat{s} \leq qg_0 + qs^*.
\end{equation}
%holds if
%\begin{equation}\label{lemmaonlinecondition}
%\hat{w}(\tau_0, \tau_1) > w^*(\tau_0, \tau_1)
%\end{equation}
\end{lemma}
\emph{Proof:}
Based on the definition, we have following two inequalities
\begin{subequations}
\begin{align}
\label{lemmaproof1}
\frac{\hat{w}(\tau_0, \tau_1)}{\tau_1 - \tau_0} \leq \sum_{i \in \mathcal{I}(\tau_0)} U_i, \\
\label{lemmaproof2}
\frac{w^*(\tau_0, \tau_1)}{\tau_1 - \tau_0} \leq \sum_{i \in \mathcal{I}(\tau_0)} U_i,
\end{align}
\end{subequations}
which hold because all the PEVs with deadlines in $[\tau_0, \tau_1]$ must park in the station at current time $\tau_0$ such that $\sum_{i \in \mathcal{I}(\tau_0)} U_i$ is larger or equal to $\sum_{i \in \mathcal{I}(t)} U_i$ for $t \in (\tau_0, \tau_1]$.
Due to the setting of $\hat{s}$ in our online algorithm, either the inequality
\begin{equation}\label{lemmaproof3}
q\frac{\hat{w}(\tau_0,\tau_1)}{\tau_1 - \tau_0} \leq \hat{s} < \sum_{i \in \mathcal{I}(\tau_0)} U_i
\end{equation}
or
\begin{equation}\label{lemmaproof4}
\hat{s} = \sum_{i \in \mathcal{I}(\tau_0)}U_i \leq q\frac{\hat{w}(\tau_0,\tau_1)}{\tau_1 - \tau_0}
\end{equation}
holds. Similarly, for optimal total charging rate $s^*$ in offline algorithm, the inequality
\begin{equation}\label{lemmaproof5}
\frac{w^*(\tau_0, \tau_1)}{\tau_1 - \tau_0} \leq s^* \leq \sum_{i \in \mathcal{I}(\tau_0)} U_i
\end{equation}
holds since $w^*(\tau_0, t)$ does not include the demand of the future coming PEVs while $s^*$ dose. From the definition of $g_k$, we get that
\begin{equation}\label{lemmaproof:g0}
\begin{aligned}
g_0 = \max \Big \{ 0, \min \{ \frac{\hat{w}(\tau_0,\tau_1)}{\tau_1 - \tau_0}, \frac{1}{q} \sum_{i \in \mathcal{I}(\tau_0)}U_i \} - \\
\min \{ \frac{w^*(\tau_0,\tau_1)}{\tau_1 - \tau_0}, \sum_{i \in \mathcal{I}(\tau_0)}U_i \} \Big \}
\end{aligned}
\end{equation}
To further reduce $g_0$, we need to discuss the following four cases.

\textit{Case 1: } If
\begin{equation}
q \frac{\hat{w}(\tau_0, \tau_1)}{\tau_1 - \tau_0} \geq \sum_{i \in \mathcal{I}(\tau_0)}U_i \text{~and~} \frac{w^*(\tau_0, \tau_1)}{\tau_1 - \tau_0} = \sum_{i \in \mathcal{I}(\tau_0)}U_i,
\end{equation}
then from (\ref{lemmaproof4}) (\ref{lemmaproof5}) we get
\begin{equation}
\hat{s} = s^* = \sum_{i \in \mathcal{I}(\tau_0)}U_i
\end{equation}
and
\begin{equation}\label{lemmaproof_g1}
\begin{aligned}
g_0 = \max \Big \{ 0, \frac{1}{q} \sum_{i \in \mathcal{I}(\tau_0)}U_i - \sum_{i \in \mathcal{I}(\tau_0)}U_i \Big\}
= 0.
\end{aligned}
\end{equation}
Hence,
\begin{equation}
qg_0 = 0 \leq \hat{s} = \sum_{i \in \mathcal{I}(\tau_0)}U_i \leq q \sum_{i \in \mathcal{I}(\tau_0)}U_i = qg_0 + qs^*.
\end{equation}

\textit{Case 2: } If
\begin{equation}
q \frac{\hat{w}(\tau_0, \tau_1)}{\tau_1 - \tau_0} <\sum_{i \in \mathcal{I}(\tau_0)}U_i \text{~and~} \frac{w^*(\tau_0, \tau_1)}{\tau_1 - \tau_0} = \sum_{i \in \mathcal{I}(\tau_0)}U_i,
\end{equation}
then from (\ref{lemmaproof3}) (\ref{lemmaproof5}) we get
\begin{equation}
\hat{s} \leq s^* = \sum_{i \in \mathcal{I}(\tau_0)}U_i
\end{equation}
and
\begin{equation}\label{lemmaproof_g2}
\begin{aligned}
g_0 = \max \Big \{ 0, \frac{ \hat{w}(\tau_0, \tau_1)}{\tau_1 - \tau_0} - \sum_{i \in \mathcal{I}(\tau_0)}U_i \Big\}
= 0.
\end{aligned}
\end{equation}
Hence,
\begin{equation}
qg_0 = 0 \leq \hat{s} \leq q \sum_{i \in \mathcal{I}(\tau_0)}U_i = qg_0 + qs^*.
\end{equation}

\textit{Case 3: } If
\begin{equation}
q \frac{\hat{w}(\tau_0, \tau_1)}{\tau_1 - \tau_0} \geq \sum_{i \in \mathcal{I}(\tau_0)}U_i \text{~and~} \frac{w^*(\tau_0, \tau_1)}{\tau_1 - \tau_0} < \sum_{i \in \mathcal{I}(\tau_0)}U_i,
\end{equation}
then from (\ref{lemmaproof4}) (\ref{lemmaproof5}) we get
\begin{equation}\label{lemmaproof6}
s^* \leq \hat{s} = \sum_{i \in \mathcal{I}(\tau_0)}U_i
\end{equation}
and
\begin{equation}
\begin{aligned}
g_0 = \max \Big \{ 0, \frac{1}{q} \sum_{i \in \mathcal{I}(\tau_0)}U_i - \frac{ w^*(\tau_0, \tau_1)}{\tau_1 - \tau_0} \Big\}.
\end{aligned}
\end{equation}
If
\begin{equation}
 \frac{1}{q} \sum_{i \in \mathcal{I}(\tau_0)}U_i \leq \frac{ w^*(\tau_0, \tau_1)}{\tau_1 - \tau_0},
\end{equation}
then we get $g_0 = 0$, and
\begin{equation}
qg_0 = 0 \leq \hat{s} = \sum_{i \in \mathcal{I}(\tau_0)}U_i \leq q \frac{ w^*(\tau_0, \tau_1)}{\tau_1 - \tau_0}\leq qs^* = qg_0 + qs^*.
\end{equation}
If
\begin{equation}
 \frac{1}{q} \sum_{i \in \mathcal{I}(\tau_0)}U_i > \frac{ w^*(\tau_0, \tau_1)}{\tau_1 - \tau_0},
\end{equation}
then we get
\begin{equation}
g_0 = \frac{1}{q} \sum_{i \in \mathcal{I}(\tau_0)}U_i - \frac{ w^*(\tau_0, \tau_1)}{\tau_1 - \tau_0}.
\end{equation}
Hence, from (\ref{lemmaproof6}) we have
\begin{subequations}
\begin{align}
&qg_0 = \sum_{i \in \mathcal{I}(\tau_0)}U_i - q\frac{ w^*(\tau_0, \tau_1)}{\tau_1 - \tau_0} \leq \hat{s},
\end{align}
\end{subequations}
and from (\ref{lemmaproof5}) we have
\begin{subequations}
\begin{align}
&qg_0 + qs^* =  \sum_{i \in \mathcal{I}(\tau_0)}U_i - q\frac{ w^*(\tau_0, \tau_1)}{\tau_1 - \tau_0} + qs^* \geq \sum_{i \in \mathcal{I}(\tau_0)}U_i = \hat{s}.
\end{align}
\end{subequations}
Since (\ref{onlineratio5}) holds for both cases, we see that (\ref{onlineratio5}) holds in Case 3.

\textit{Case 4: } If
\begin{equation}
q \frac{\hat{w}(\tau_0, \tau_1)}{\tau_1 - \tau_0} < \sum_{i \in \mathcal{I}(\tau_0)}U_i \text{~and~} \frac{w^*(\tau_0, \tau_1)}{\tau_1 - \tau_0} < \sum_{i \in \mathcal{I}(\tau_0)}U_i,
\end{equation}
we get
\begin{equation}\label{lemmaproof10}
g_0 = \max \Big \{ 0, \frac{ \hat{w}(\tau_0, \tau_1)}{\tau_1 - \tau_0} -  \frac{ w^*(\tau_0, \tau_1)}{\tau_1 - \tau_0} \Big\}.
\end{equation}
When $\hat{w}(\tau_0, \tau_1) \geq w^*(\tau_0, \tau_1)$, (\ref{lemmaproof:g0}) is reduced to
\begin{equation}
g_0 = \frac{ \hat{w}(\tau_0, \tau_1)}{\tau_1 - \tau_0} -  \frac{ w^*(\tau_0, \tau_1)}{\tau_1 - \tau_0}.
\end{equation}
Recall that $\hat{s} = q s^{OA}$ where $s^{OA}$ is the total charging rate of OA algorithm given the current demand $\hat{w}\left(\tau_0,\tau_1\right)$, and $s^*$ is the charging rate given $w^*\left(\tau_0,\tau_1\right)$ and possible future arrivals of other PEVs. Notice that both $\hat{w}(\tau_0, \tau_1)$ and $w^*(\tau_0, \tau_1)$ do not include the charging demand of future coming PEVs, and the difference between $s^{OA}$ and $\hat{w}(\tau_0, \tau_1)/(\tau_1 - \tau_0)$ is only resulted from the bounds of current PEVs
, while the difference between $s^*$ and $w^*(\tau_0, \tau_1)/(\tau_1 - \tau_0)$ is due to the bounds of current PEVs as well as the possible heavy load of future coming PEVs. Therefore,
\begin{equation}\label{lemmaproof9}
s^{OA} - \frac{\hat{w}(\tau_0, \tau_1)}{\tau_1 - \tau_0} \leq s^* - \frac{w^*(\tau_0, \tau_1)}{\tau_1 - \tau_0}
\end{equation}
As $\hat{s} = q s^{OA}$, we have
\begin{equation}\label{lemmaproof9a}
\hat{s} - q\frac{\hat{w}(\tau_0, \tau_1)}{\tau_1 - \tau_0} \leq qs^* - q\frac{w^*(\tau_0, \tau_1)}{\tau_1 - \tau_0}.
\end{equation}
Hence we get the following inequalities:
\begin{subequations}
\begin{align}
\label{lemmaproof7}
&qg_0 = q\frac{ \hat{w}(\tau_0, \tau_1)}{\tau_1 - \tau_0} -  q\frac{ w^*(\tau_0, \tau_1)}{\tau_1 - \tau_0} \leq \hat{s}, \\
\label{lemmaproof8}
&qg_0 + qs^* \geq  (q \frac{ \hat{w}(\tau_0, \tau_1)}{\tau_1 - \tau_0} - q\frac{ w^*(\tau_0, \tau_1)}{\tau_1 - \tau_0})  \nonumber\\
&+ (\hat{s} - q\frac{\hat{w}(\tau_0, \tau_1)}{\tau_1 - \tau_0} + q \frac{ w^*(\tau_0, \tau_1)}{\tau_1 - \tau_0}) = \hat{s},
\end{align}
\end{subequations}
where the last inequality of (\ref{lemmaproof7}) and the first inequality of (\ref{lemmaproof8}) are derived from (\ref{lemmaproof3}) and (\ref{lemmaproof9a}) respectively.
For the case that $\hat{w}(\tau_0, \tau_1) < w^*(\tau_0, \tau_1)$, we have $g_0 = 0$ and $\hat{s} \leq q s^*$ by adding on left hand side of (\ref{lemmaproof9a}) $q\hat{w}(\tau_0, \tau_1)/(\tau_1 -\tau_0)$ and right hand side $qw^*(\tau_0, \tau_1)/(\tau_1 -\tau_0)$. Therefore, (\ref{onlineratio5}) holds in case 4.

Finally, inequality (\ref{onlineratio5}) holds in all the four cases. This completes the proof.
\hfill $\blacksquare$

In the following Theorem \ref{theormonline}, we derive the competitive ratio of ORCHARD.
\begin{theorem}\label{theormonline}
ORCHARD is $2.39$-competitive by setting $q = 1.46$.
\end{theorem}
\emph{Proof:} We can derive from (\ref{onlineratio6}) that
\begin{equation}\label{ratioproof1}
%\small
\begin{aligned}
\frac{\mathrm d \Phi}{\mathrm d \tau_0}= & \beta_1 \cdot a \sum_{k = 0}^{\infty}\frac{\mathrm d \left[(\tau_{k+1} - \tau_k) g_k \right]}{\mathrm d \tau_0} \\
& + \beta_2 \cdot b \sum_{k = 0}^{\infty}\frac{\mathrm d \left[(\tau_{k+1} - \tau_k) g_k^2 \right]}{\mathrm d \tau_0}.
\end{aligned}
\end{equation}
When $\hat{w}(\tau_0, \tau_1)< w^*(\tau_0, \tau_1)$, we divide into following four cases to prove that $g_0 = 0$, where $\tau_1$ is infinity.

\begin{enumerate}
  \item If
\begin{equation}
q \frac{\hat{w}(\tau_0, \tau_1)}{\tau_1 - \tau_0} \geq \sum_{i \in \mathcal{I}(\tau_0)}U_i \text{~and~} \frac{w^*(\tau_0, \tau_1)}{\tau_1 - \tau_0} = \sum_{i \in \mathcal{I}(\tau_0)}U_i,
\end{equation}
then (\ref{lemmaproof_g1}) implies that $g_0 = 0$.
  \item If
\begin{equation}
q \frac{\hat{w}(\tau_0, \tau_1)}{\tau_1 - \tau_0} <\sum_{i \in \mathcal{I}(\tau_0)}U_i \text{~and~} \frac{w^*(\tau_0, \tau_1)}{\tau_1 - \tau_0} = \sum_{i \in \mathcal{I}(\tau_0)}U_i,
\end{equation}
then (\ref{lemmaproof_g2}) implies that $g_0 = 0$.
  \item If
\begin{equation}
q \frac{\hat{w}(\tau_0, \tau_1)}{\tau_1 - \tau_0} \geq \sum_{i \in \mathcal{I}(\tau_0)}U_i \text{~and~} \frac{w^*(\tau_0, \tau_1)}{\tau_1 - \tau_0} < \sum_{i \in \mathcal{I}(\tau_0)}U_i,
\end{equation}
then
\begin{equation}
q \frac{w^*(\tau_0, \tau_1)}{\tau_1 - \tau_0} \geq q \frac{\hat{w}(\tau_0, \tau_1)}{\tau_1 - \tau_0} \geq \sum_{i \in \mathcal{I}(\tau_0)}U_i.
\end{equation}
Hence,
\begin{equation}
\begin{aligned}
g_0 &=& \max \Big \{ 0, \frac{1}{q} \sum_{i \in \mathcal{I}(\tau_0)}U_i - \frac{ w^*(\tau_0, \tau_1)}{\tau_1 - \tau_0} \Big\} = 0.
\end{aligned}
\end{equation}
  \item If
\begin{equation}
q \frac{\hat{w}(\tau_0, \tau_1)}{\tau_1 - \tau_0} < \sum_{i \in \mathcal{I}(\tau_0)}U_i \text{~and~} \frac{w^*(\tau_0, \tau_1)}{\tau_1 - \tau_0} < \sum_{i \in \mathcal{I}(\tau_0)}U_i,
\end{equation}
then
\begin{equation}\label{lemmaproof10}
g_0 = \max \Big \{ 0, \frac{ \hat{w}(\tau_0, \tau_1)}{\tau_1 - \tau_0} -  \frac{ w^*(\tau_0, \tau_1)}{\tau_1 - \tau_0} \Big\} = 0.
\end{equation}
\end{enumerate}

Hence, $g_0 = 0$ holds when $\hat{w}(\tau_0, \tau_1)< w^*(\tau_0, \tau_1)$. Then $\mathrm d\Phi / \mathrm d \tau_0$ remains zero and $\hat{s} \leq qs^*$ by Lemma \ref{lemmaonline}. Then, (\ref{onlineratio1}) always holds by letting $ q^2 \leq c$. Therefore, we only consider the case that $\hat{w}(\tau_0, \tau_1)\geq w^*(\tau_0, \tau_1)$ with $q^2 \leq c$.
For the speed scaling problem in \cite{bansal2009speed}, since there is no constraint of scheduling rate for each individual job, both the online and offline algorithm can always have a solution that only schedule one job that the load intensity gap varies only in at most two time intervals. However, in our problem, since for any PEV, its charging rate can not exceed the maximum charging rate, this leads to that the scheduler should at least charging one PEV at time $\tau_0$. Then we should compute the differential of intensity gap for all intervals and then combine them together.
For the time interval $[\tau_0, \tau_1]$, we have
\begin{equation}\label{ratioproof2a}
%\small
\begin{aligned}
     &\frac{\mathrm d (\tau_1 - \tau_0)g_0} {\mathrm d \tau_0}  \\
    =& (\tau_1 - \tau_0)\frac{(\tau_1 - \tau_0) \frac{\mathrm{d} d(\tau_0, \tau_1)}{\mathrm{d} \tau_0} + d(\tau_0, \tau_1) }{(\tau_1 - \tau_0)^2} - g_0^2
    = \frac{\mathrm d  d(\tau_0,\tau_1)}{\mathrm d \tau_0}
\end{aligned}
\end{equation}
and
\begin{equation}\label{ratioproof2}
%\small
\begin{aligned}
     &\frac{\mathrm d (\tau_1 - \tau_0)g_0^2} {\mathrm d \tau_0}  \\
    =& 2g_0(\tau_1 - \tau_0)\frac{(\tau_1 - \tau_0) \frac{\mathrm{d} d(\tau_0, \tau_1)}{\mathrm{d} \tau_0} + d(\tau_0, \tau_1) }{(\tau_1 - \tau_0)^2} - g_0^2 \\
    =& 2 g_0\frac{\mathrm d  d(\tau_0,\tau_1)}{\mathrm d \tau_0} + g_0^2.
\end{aligned}
\end{equation}
For the time interval $(\tau_k, \tau_{k+1}], k = 1, 2, \ldots$, we have
\begin{equation}\label{ratioproof3a}
%\small
\begin{aligned}
    \frac{\mathrm d ((\tau_{k+1} - \tau_k)g_k)}{\mathrm d \tau_0}
    = \frac{\mathrm d  d(\tau_k,\tau_{k+1})}{\mathrm d \tau_0}
    < \frac{\mathrm d  d(\tau_k,\tau_{k+1})}{\mathrm d \tau_0},
\end{aligned}
\end{equation}
and
\begin{equation}\label{ratioproof3}
%\small
\begin{aligned}
    \frac{\mathrm d ((\tau_{k+1} - \tau_k)g_k^2)}{\mathrm d \tau_0}
    = 2 g_k \frac{\mathrm d  d(\tau_k,\tau_{k+1})}{\mathrm d \tau_0}
    < 2 g_0 \frac{\mathrm d  d(\tau_k,\tau_{k+1})}{\mathrm d \tau_0},
\end{aligned}
\end{equation}
where the last inequality holds because $g_k<g_0$, $\forall k>0$. Summing up (\ref{ratioproof2a})(\ref{ratioproof2})(\ref{ratioproof3a}) and (\ref{ratioproof3}), $ \mathrm{d} \Phi /\mathrm{d} \tau_0$ is upper bounded by
\begin{equation}\label{ratioproof4}
%\small
\begin{aligned}
&\beta_1 a \left( \sum_{k=0}^\infty \frac{\mathrm d  d(\tau_k,\tau_{k+1})}{\mathrm d \tau_0} \right) + \beta_2 b \left( 2 g_0\sum_{k=0}^\infty \frac{\mathrm d  d(\tau_k,\tau_{k+1})}{\mathrm d \tau_0} + g_0^2\right)\\
=& \beta_1 a(-\hat{s} + s^*) + \beta_2 b \left( 2 g_0(-\hat{s}+ s^*)  + g_0^2\right).
\end{aligned}
\end{equation}
Then, to prove (\ref{onlineratio1}), it suffices to show that the following inequality holds, where
\begin{equation}\label{ratioproof5}
\begin{aligned}
    &(a\hat{s} + b(\hat{s})^2) + (\beta_1 a(-\hat{s} + s^*) + \beta_2 b ( 2 g_0(-\hat{s}+ s^*)  + g_0^2)) \\
    &- c(as^* + b(s^*)^2) \leq 0.
\end{aligned}
\end{equation}
And it is also suffices to show that the following two inequalities hold, where
\begin{subequations}\label{ratioproof6}
\begin{align}
    \label{onlineratio7} &a\hat{s}  + \beta_1 a(-\hat{s} + s^*) - c\cdot as^* \leq 0 \\
    \label{onlineratio8} & b(\hat{s})^2 + \beta_2 b ( 2 g_0(-\hat{s}+ s^*)  + g_0^2) - c\cdot b(s^*)^2 \leq 0.
\end{align}
\end{subequations}
Notice that the LHS of (\ref{onlineratio7}) is a linear function of $\hat{s}$, it therefore suffices to show that (\ref{onlineratio7}) holds for all $s^* \geq 0$ and $g_0 \geq 0$ when $\hat{s} = qg_0$ and $\hat{s} = q(s^*+g_0)$, i.e.,
\begin{subequations} \label{onlineratio9}
\begin{align}
   & (1 - \beta_1)qg_0 + (\beta_1 - c)s^* \leq 0 \\
   & (1 - \beta_1)(qg_0 + qs^*) + (\beta_1 - c)s^* \leq 0.
\end{align}
\end{subequations}
Since $c \geq 1, q \geq 1$, by setting $\beta_1 = 1$, (\ref{onlineratio9}) holds for all $s^* \geq 0$ and $g_0 \geq 0$.
Similarly, since the LHS of (\ref{onlineratio8}) is a convex function of $\hat{s}$, it therefore suffices to show that (\ref{onlineratio8}) holds for all $s^* \geq 0$ and $g_0 \geq 0$ when $\hat{s} = qg_0$ and $\hat{s} = q(s^*+g_0)$. To obtain the lowest competitive ratio, we need to determine the values of $q (1 \leq q^2 \leq c)$ and $\beta_2$ that minimize $c$. This can be achieved by using the numerical method in \cite{bansal2009improved}. We do not present the detailed steps but only the numerical results. That is, the optimal parameters are $q = 1.46$ and $\beta_2 = 2.7$, where the lowest competitive ratio is $2.39$. \hfill $\blacksquare$

%In fact, the derived competitive ratio coincides with that of the processor speed scaling problem studied in \cite{bansal2009improved}, which is the best possible competitive ratio so far. Note that PEV charging problem is more generalized than speed scaling on both objective and constrains. Hence, ORCHARD achieves the best known competitive ratio for online PEV charging problem.

\section{A Low Complexity Solution Algorithm to Problem (\ref{model6}) and (\ref{model16})}\label{sec:offline algorithm}
The major complexity of Algorithm \ref{onlinealgorithm} lies in the computation involved in solving Problem (\ref{model16}) every time when a PEV arrives or finishes charging. By exploring the special structure of the optimal solution, we propose in this section a low-complexity solution algorithm to solve problem (\ref{model16}). Notice that Problem (\ref{model16}) and the offline optimization problem (\ref{model6}) have exactly the same structure. Both of them are to minimize a convex and additive objective function over a polyhedron. Thus, the algorithm proposed here can also apply to (\ref{model16}). The proposed algorithm is shown to have a much lower computational complexity than generic convex optimization algorithms, such as interior point method.

\subsection{KKT Optimality Conditions}\label{sec:kkt}
The KKT conditions to the convex problem (\ref{model6}) are
\begin{subequations}
%\small
\begin{align}
\label{kkt1}
  &a + 2b \sum_{j \in \mathcal{I}(k)}x^*_{jk}
   - \lambda_i + \nu_{ik} - \omega_{ik} = 0,
   i = 1,\ldots,N, k \in \mathcal{J}(i). \\
\label{kkt2}
  &\lambda_i(D_i - \sum_{k \in \mathcal{J}(i)} x^*_{ik}) = 0, i = 1,\ldots,N. \\
\label{kkt3}
  &\omega_{ik}x^*_{ik} = 0, i = 1,\ldots,N, k \in \mathcal{J}(i). \\
\label{kkt4}
  &\nu_{ik}(x^*_{ik} - U_i) = 0, i = 1,\ldots,N, k \in \mathcal{J}(i).
\end{align}
\end{subequations}
where $\lambda, \omega$ and $\nu$ are the non-negative optimal Lagrangian multipliers to the respective constraints. We separate our analysis into the following three cases:
\begin{enumerate}
  \item If $x^*_{ik_1}\in \left(0,U_i\right)$  for a particular PEV $i$ in a time interval $k_1 \in \mathcal{J}(i)$, then, by complementary slackness, we have $\nu_{ik_1} = w_{ik_1} = 0$. From (\ref{kkt1}), $s_{k_1} = \sum_{j \in \mathcal{I}(k_1)}x_{jk_1}= (\lambda_i -a)/2b$.
  \item If $x^*_{ik_2} = 0$ for PEV $i$ during a time interval $k_2 \in \mathcal{J}(i)$, we can infer from (\ref{kkt3}) and (\ref{kkt4}) that $\omega_{ik_2} > 0$ and $\nu_{ik_2} = 0$. Then, $s_{k_2} = \sum_{j \in \mathcal{I}(k_2)}x^*_{jk_2}= \left(\lambda_i -a\right)/2b + \omega_{ik_2}/2b$.
  \item Similarly, if $x^*_{ik_3} = U_i$  for PEV $i$ in interval $k_3 \in \mathcal{J}(i)$, then, we have $s_{k_3} = \sum_{j \in \mathcal{I}(k_3)}x^*_{jk_3}= \left(\lambda_i -a\right)/2b - \nu_{ik_3}/2b$.
\end{enumerate}
From the above discussions, we can conclude that the necessary and sufficient conditions for the optimal total charging rate as follows:
\begin{enumerate}
  \item $s^*_k$ is the same for a set of intervals as long as there exists a PEV $i$ that parks through this set of intervals with $x^*_{ik} \in (0,U_i)$.
  \item If $x^*_{ik}=0$  for a PEV $i$ during an interval $k$ that it parks in, then, $s^*_k$ in that interval is no smaller than that of the other interval $k' \in \mathcal{J}(i)$ during which $x^*_{ik'} \in (0,U_i]$.
  \item If $x^*_{ik}=U_i$  for PEV $i$ during an interval $k$, then, $s^*_k$ is no larger than that of the other interval $k' \in \mathcal{J}(i)$ whose charging rate $x^*_{ik'} \in [0,U_i)$.
\end{enumerate}

The above conditions can be intuitively understood as follows. Due to the convexity the objective function, the optimal solution to (\ref{model6}) always tries to balance the total charging rate $s_k$ among different $k'$s. For example, if there are intervals $k_1$ and $k_2$ with $s^*_{k_1}>s^*_{k_2}$, and a PEV $i$ such that $x^*_{ik_1}>0$ and $x^*_{ik_2}=0$, then we can always shift the charging rate of PEV $i$ from interval $k_1$ to $k_2$ to decrease the total cost. In other words, whenever possible, the charging rate should be shifted from intervals with higher total charging rates to the ones with lower total charging rates until the limits, i.e., $0$ and $U_i$'s, have been reached. Based on these conditions, we will present a low-complexity solution algorithm in the next subsection.

\subsection{Algorithm Description}
\begin{figure}
        \centering
        \includegraphics[width=8cm,height=4cm]{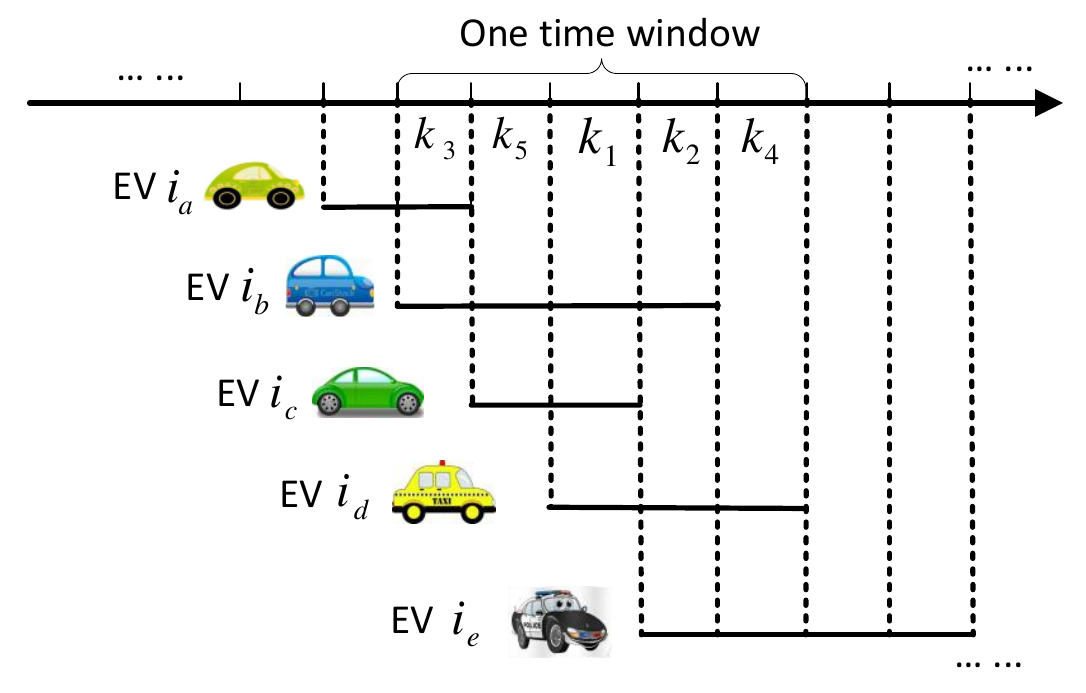}
        \caption{Illustration of one time window}
        \label{fig:offlinealgorithm}
        %\vspace{-0.6cm}
\end{figure}

From the analysis of KKT optimality conditions, one should manage to balance the charging load among all intervals under the constraints of each individual PEV's charging profiles. In this subsection, we present a charging rate allocation algorithm to achieve the objective of ``load balancing". The optimality and complexity of the proposed algorithm will be discussed in the next subsection.

Intuitively, one should shift the demand from ``heavily loaded'' intervals to the others. To do this, we first introduce the concept of \emph{intensity} of an interval $k$, denoted by $\rho_k$,
%defined as the upper bound of the total charging rate of an interval,
to quantify the heaviness of the load in the interval.
Specifically, $\rho_k$ is defined
as the upper bound of the total charging rate of an interval, and is given by
\begin{equation}
%\small
\rho_k = \sum_{i \in \mathcal{I}(k)} \min \left \{U_i, \frac{D_i}{\delta_k} \right \}.
\end{equation}
This is because the charging rate of each PEV $i$ in the interval $k$ will not exceed the minimum between the charging rate bound $U_i$ and the $D_i / \delta_k$, i.e. PEV $i$ only charges in the interval $k$. The basic idea of the proposed algorithm is to shift the demand of a set of intervals with high intensities to the others with lower intensities. Notice that the demand of an interval $k_1$ can only be transferred to its neighboring interval $k_2$ such that $k_1\in J(i)$ and $k_2\in J(i)$ hold for some PEV $i$. Therefore, we need consider both the intensities of an interval set and their neighboring intervals to make the decision on ``load balancing''.

From the above discussion, we take into consideration a set of consecutive intervals, referred to as a ``time window'', starting from the arrival time of a PEV to the departure time of one, probably another PEV. If there are $N$ PEVs, the maximum number of time windows is $N^2$. Within a tagged time window, we select a set of intervals of the highest intensities as the candidate interval set from which the load is to be transferred to the other intervals in the time window. In practice, we first consider the single interval with the highest intensity, then the top two intervals, top three intervals, etc. That is, for each time window, sort the intervals in descending order according to $\rho$. The index is denoted by  $k_1, k_2, \ldots$, as illustrated in Fig.~\ref{fig:offlinealgorithm}.
%Let $\mathcal{K} = \{k_1, k_2, \ldots\}$ be a subset of time intervals with highest $\rho_k$ in the certain time window.

Evidently, a time window consisting of $K'$ intervals contains $K'$ such interval sets. For example, there are $5$ interval sets in the time window shown in Fig.~\ref{fig:offlinealgorithm}. We denote the interval sets obtained from all the time windows in the entire duration $T$ as $\mathcal{K}_1$, $\mathcal{K}_2$, $\cdots$.  Then, the following iterative algorithm determines the load transfer operation of intervals as well as the charging rate schedule of all PEVs.

%The computation steps are given in Algorithm \ref{algorithm:offline} and explained as follows:
%The computation steps of the interative algorithm are explained as follows:

\textit{Step 1: }
For each interval set $\mathcal{K}$, we first compute the residual demand of PEV $i$
on $\mathcal{K}$.
The residual demand of PEV $i$ on $\mathcal{K}$, denoted by $D_{i}(\mathcal{K})$, is calculated by letting PEV $i$ be charged at the upper bound $U_i$ on its parking intervals non-overlapped with $\mathcal{K}$.
That is
\begin{equation}
%\small
\label{residualD}
D_{i}(\mathcal{K}) = D_i - U_i \sum_{k \in \mathcal{J}(i) \setminus ( \mathcal{J}(i) \cap \mathcal{K} ) } \delta_k .
\end{equation}
%For each PEV $i$, $\rho_k$ at $ k \in \mathcal{J}(i) \cap \mathcal{K}$ must be higher than that at other parking intervals for each PEV $i$ so that intuitively we should try best to transfer the charging demand from interval $k \in \mathcal{J}(i) \cap \mathcal{K}$ to other parking intervals. The extreme case is that if it is charged at $u_i$ on other parking interval set $\mathcal{J}(i) - \mathcal{J}(i) \cap \mathcal{K}$.
The intuition is to transfer as much as possible the charging demand from intervals with high intensities to its neighboring intervals.
%If we charge the PEVs on parking other intervals at their charging bounds and then balance the whole residual demand over the  time intervals of $\mathcal{K}$, that is indeed the equally balanced total charging rate of interval set $\mathcal{K}$, i.e.,
Then, we can calculate the total charging rate of the interval set $\mathcal{K}$ by balancing the residual demand over all the intervals in $\mathcal{K}$, i.e.,
\begin{equation}\label{residual_intensity}
%\small
s = \frac{\sum_{k \in \mathcal{K}} (\sum_{i \in \mathcal{I}(k)}(\max\{0, D_{i}(\mathcal{K}) \}) + \hat{s}_k \delta_k )}{\sum_{k \in \mathcal{K}} \delta_k},
\end{equation}
where $\hat{s}_k$ is the charging rate scheduled in previous iterations at the interval and initially set to be $0$.

  \textit{Step 2: }
  Find the interval set $\mathcal{K}^*$ with the highest total charging rate $s^*$. Then the optimal total charging rate of interval in $ \mathcal{K}^*$ is set to be $s^*$, i.e.,
  \begin{equation} \label{totalchargingrate}
  s_k^*= s^*, \forall k \in \mathcal{K}^*.
  \end{equation}
  We denote $\mathcal{I}^*$ by the set of PEVs of which the residual demand $D_i(\mathcal{K}^*)$ is non-negative, $\Delta^*$ by the total length of the intervals in the set $\mathcal{K}^*$, i.e., $\Delta^*= \sum_{k \in \mathcal{K}^*}\delta_k$.
  %Then, we should charge the interval set $\mathcal{K}^*$ first. That is, the optimal total charging rate of interval $k \in \mathcal{K}^*$ is
%  \begin{equation}\label{optimalchargingload}
%  \small
%  s_k^*= s^*, \forall k \in \mathcal{K}^*.
%  \end{equation}
  For each PEV $i \in \mathcal{I}^*$, we schedule the charging rate as
  \begin{equation}\label{chargingrate}
  %\small
  \begin{aligned}
  x_{ik}^*=
  \begin{cases}
U_i - \frac{ (U_i \Delta^*- D_{i}(\mathcal{K}^*)) (\sum_i U_i - s_k^*) } {\sum_i(U_i \Delta^* - D_{i}(\mathcal{K}^*))}, &   k\in \mathcal{K}^*,\\
U_i,   &   k\in \mathcal{J}(i) \setminus \mathcal{K}^*.\\
\end{cases}
\end{aligned}
\end{equation}
It is easy to verify that $\sum_{k\in \mathcal{K}^*} x_{ik}^* = s_k^*$ for $k \in \mathcal{K}^*$.
Note that PEV $i \in \mathcal{I}^*$ has finished scheduled charging rate and won't be considered in the next iterations. However, its charging rate at interval $k \in \mathcal{J}(i) \setminus \mathcal{K}^*$ is fixed as $U_i$ that should be considered as one component of load intensity of interval $k \in \mathcal{J}(i) \setminus \mathcal{K}^*$ in the next iteration.
%We use $\hat{s}_k$ to denote this event and update $\hat{s}_k$ in (\ref{residual_intensity}) for $k \in \mathcal{J}(i) \setminus \mathcal{K}^*$ as
We use $\hat{s}_k$ to denote the total rate scheduled in the interval $k \notin \mathcal{K}^*$ up to the current iteration, which is updated as.
  \begin{equation}\label{hat_s}
  %\small
  \hat{s}_k = \hat{s}_k + \sum_{i \in \mathcal{I}^* \cap \mathcal{I}(k)}U_i.
  \end{equation}
%Hence, we have scheduled the charging rates of PEV $i \in I^*$ now.
For a PEV $i \notin I^*$ whose parking intervals overlaps with $\mathcal{K}^*$,
the charging rate of its parking intervals overlapped with $\mathcal{K}^*$ is assigned to be $0$, i.e.,
  \begin{equation}\label{zerochargingrate}
  %\small
  x_{ik}^* = 0, k \in \mathcal{J}(i) \cap \mathcal{K}^*.
  \end{equation}

  \textit{Step 3: } Exclude $\mathcal{I}^*$ and $\mathcal{K}^*$ from the PEV set and interval set, and merge the remaining intervals into a new time duration. Find all the interval sets in the newly formed time windows as in Fig.~\ref{fig:offlinealgorithm}. Then, repeat from step $1$ until the charging rates of all PEVs are scheduled.

\subsection{Optimality and Complexity}\label{sec:offline analysis}
%Denote $l^*(m)$ as the optimal time window found in $m^{th}$ iteration, $\mathcal{K}_{l^*}(m)$, $\mathcal{K}_{l^*}^{n^*}(m)$, $\Delta_{l^*}(m)$ and $s_{n^*}^{l^*}(m)$ as the interval set, best sequence, the length of best sequence and the largest intensity of optimal time window $l^*$ respectively.
We first provide the following Lemma \ref{lemmaoffline} before proving the global optimality of the proposed algorithm. Denote $\mathcal{K}^*(m)$ by the interval set found in $m^{th}$ iteration, $\Delta^*(m)$ by the total length of intervals in $\mathcal{K}^*(m)$, i.e,
\begin{equation}
\Delta^*(m) = \sum_{k \in \mathcal{K}^*(m)}\delta_k,
\end{equation}
$s^*(m)$ by the highest total charging rate of interval set $\mathcal{K}^*(m)$ respectively.

\begin{lemma}\label{lemmaoffline}
In the proposed algorithm, the highest total charging rate found in $m^{th}$ iteration is no smaller than that found in $(m+1)^{th}$ iteration, i.e., $s^*(m) \geq s^*(m+1)$.
\end{lemma}

%Please refer to Lemma \ref{lemmaoffline} in \cite{tang13report} for detailed proof.
%Due to the space limit, the proof of Lemma \ref{lemmaoffline} is ignored here. Interested readers are referred to our technical report \cite{tang13report} for the detailed proof.
\emph{Proof:} We give the proof by contradiction. Actually, in $m^{th}$ iteration, one of the candidate interval sets is the union of intervals sets $\mathcal{K}^*(m)$ and $\mathcal{K}^*(m+1)$, denoted by $\mathcal{K}'$, i.e.,
\begin{equation}
\mathcal{K}' = \mathcal{K}^*(m) \cup \mathcal{K}^*(m+1).
\end{equation}
Note that the interval set $\mathcal{K}^*(m)$ and $\mathcal{K}^*(m+1)$ have no intersections, i.e.,
\begin{equation}
\mathcal{K}^*(m) \cap \mathcal{K}^*(m+1) = \emptyset,
\end{equation}
then the total residual demand of $\mathcal{K}'$ is the sum of residual demand of $\mathcal{K}^*(m)$ and $\mathcal{K}^*(m+1)$, i.e., $s^*(m)\Delta^*(m) + s^*(m+1)\Delta^*(m+1)$. Thus, the total charging rate of the interval set $\mathcal{K}'$ is the balanced residual demand over all the intervals in $\mathcal{K}'$, that is,
\begin{equation}
%\small
\begin{aligned}
  s &= \frac{ s^*(m)\Delta^*(m) + s^*(m+1)\Delta^*(m+1) }{ \Delta^*(m) + \Delta^*(m+1) } \\
    &> s^*(m).
\end{aligned}
\end{equation}
where the last inequality holds because $s^*(m) < s^*(m+1)$. Thus, it makes a contradiction with $s \leq s^*(m)$ since $s^*(m)$ is the highest total charging rate over all candidate interval sets in iteration $m$.
This completes the proof. \hfill $\blacksquare$

\begin{theorem}\label{theorem:offline optimal}
The proposed algorithm always outputs a globally optimal schedule.
\end{theorem}

\emph{Proof:} For any PEV $i$, assume that there exists interval $k_1, k_2, k_3 \in \mathcal{J}(i)$ where $x^*_{ik_1} = 0$, $x^*_{ik_2} \in (0, U_i)$ and $x^*_{ik_3} = U_i$. We separate the proof into the following three parts to match with the three cases of KKT optimality conditions:
\begin{enumerate}
  \item Interval $k_1$ must be excluded before interval $k_2$ and interval $k_3$ since when schedule $x^*_{ik_1} = 0$ from (\ref{zerochargingrate}), the considered PEV $i$ has not been scheduled that interval $k_2$ and $k_3$ should be reserved and goto next iteration. By Lemma \ref{lemmaoffline}, we have $s_{k_1}^* \geq s_{k_2}^*$ and $s_{k_1}^* \geq s_{k_3}^*$.
  \item Interval $k_2$ must be excluded before interval $k_3$ since when schedule $x^*_{ik_2}$ and $x^*_{ik_3}$ from (\ref{chargingrate}), interval $k_2$ belongs to the interval set with highest total charging rate and will be excluded in the current iteration, while interval $k_3$ should be reserved to next iteration. Similarly, by lemma \ref{lemmaoffline} we have $s_{k_2}^* \geq s_{k_3}^*$.
  \item For any other interval $k' \in \mathcal{J}(i)$ with $x^*_{ik'} \in (0, U_i)$, $s_{k'}^*$ is the same as $s_{k_2}^*$ because both $k'$ and $k_2$ belongs to the set $\mathcal{K}^*$ in the same iteration by (\ref{chargingrate}) and are assigned the same optimal total charging rate from (\ref{totalchargingrate}).
\end{enumerate}
Therefore, our algorithm satisfies $KKT$ conditions that the solution is always global optimal.\hfill $\blacksquare$

%\subsection{Complexity Analysis}\label{sec:offline prf}
Now we give a complexity analysis of the proposed algorithm. Consider the worst case where $N$ PEVs lead to $2N - 1$ intervals, $N^2$ variables and $2N^2 + N$ constrains.
%In general, we denote by $n$ the variable number, then $O(n) = O(N^2)$.
It at least excludes one interval in each outer loop that leads to at most $2N-1$ iterations. In each iteration (step 1 - step 3), there are at most $N(N+1)/2$ time windows which contains at most $N$ possible interval sets. Hence, the total number of iterations is in the order of $O(N^4)$. Since the operation complexity of intensity calculation for each sequence is $O(N)$ (we regard one addition, subtraction, multiplication and division as one operation), the upper bound of operation complexity is $O(N^5)$.
On the other hand, the generic interior point algorithm has a complexity at the order of $O(n^{3.5})$ \cite{ye97}, where $n$ is the number of variables. Note that $n=N^2$ in our problem, and thus the complexity of interior point algorithm is $O(N^7)$, which is much higher than that of the proposed algorithm.
%Compared to interior point algorithm, which has a complexity $O(n^{3.5}log(R/\epsilon))$ \cite{ye97}, the proposed the algorithm reduces the complexity by exploring the structure of the optimal solutions.
%And compared to Ellipsoid method, its arithmetic operation complexity is $O(n^4ln(2+VAR_x * R/r\epsilon))$ for general convex programming \cite{ye97}.

\section{Simulations}\label{sec:sim}
\begin{figure*}
%\vspace{0.3cm}
\centering
\subfigure[Scenario 1: light traffic]{
    \includegraphics[width=0.31\textwidth]{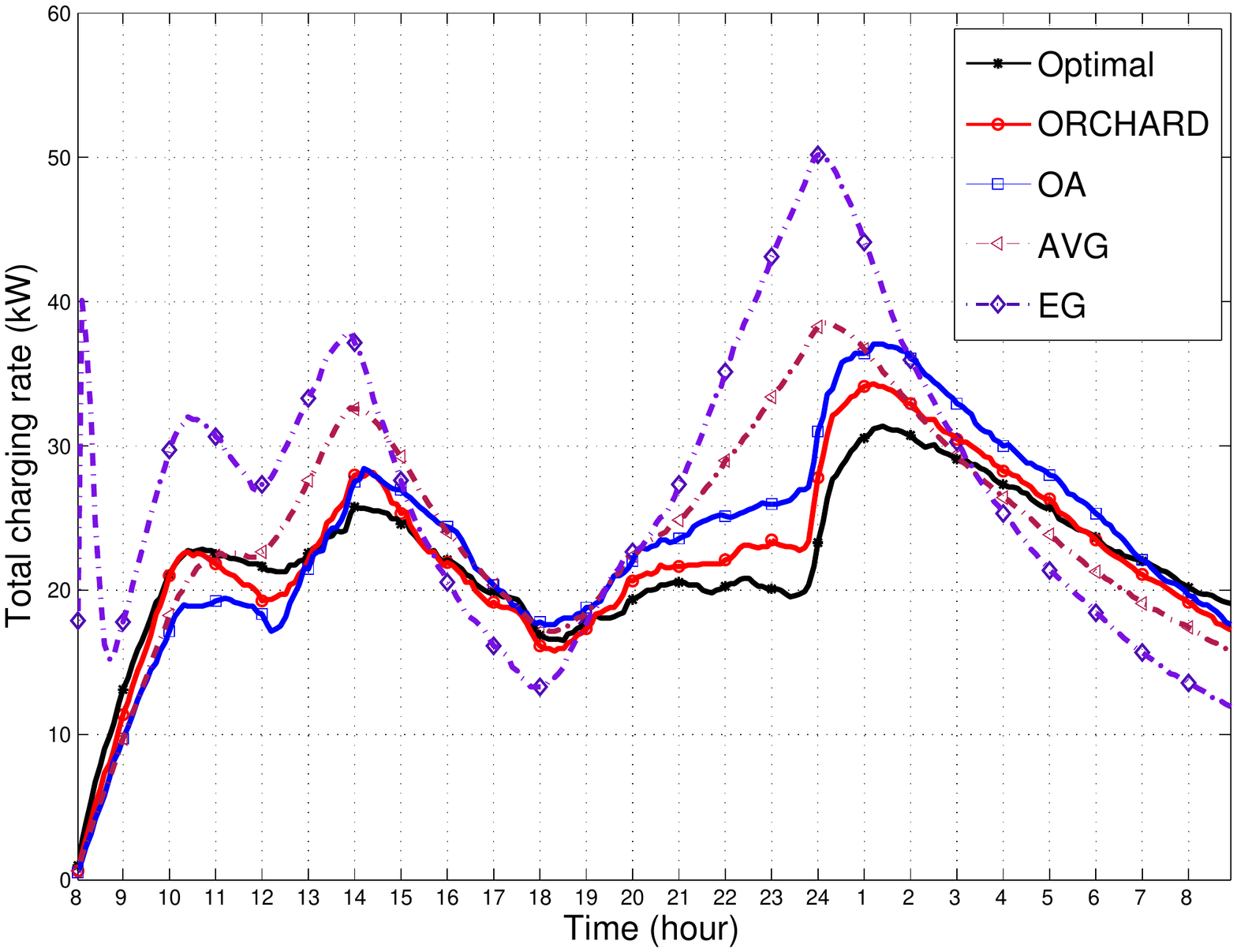}
        \label{fig:oneday_flat}
}
%\hspace{-0.05in}
\subfigure[Scenario 2: moderate traffic]{
    \includegraphics[width=0.31\textwidth]{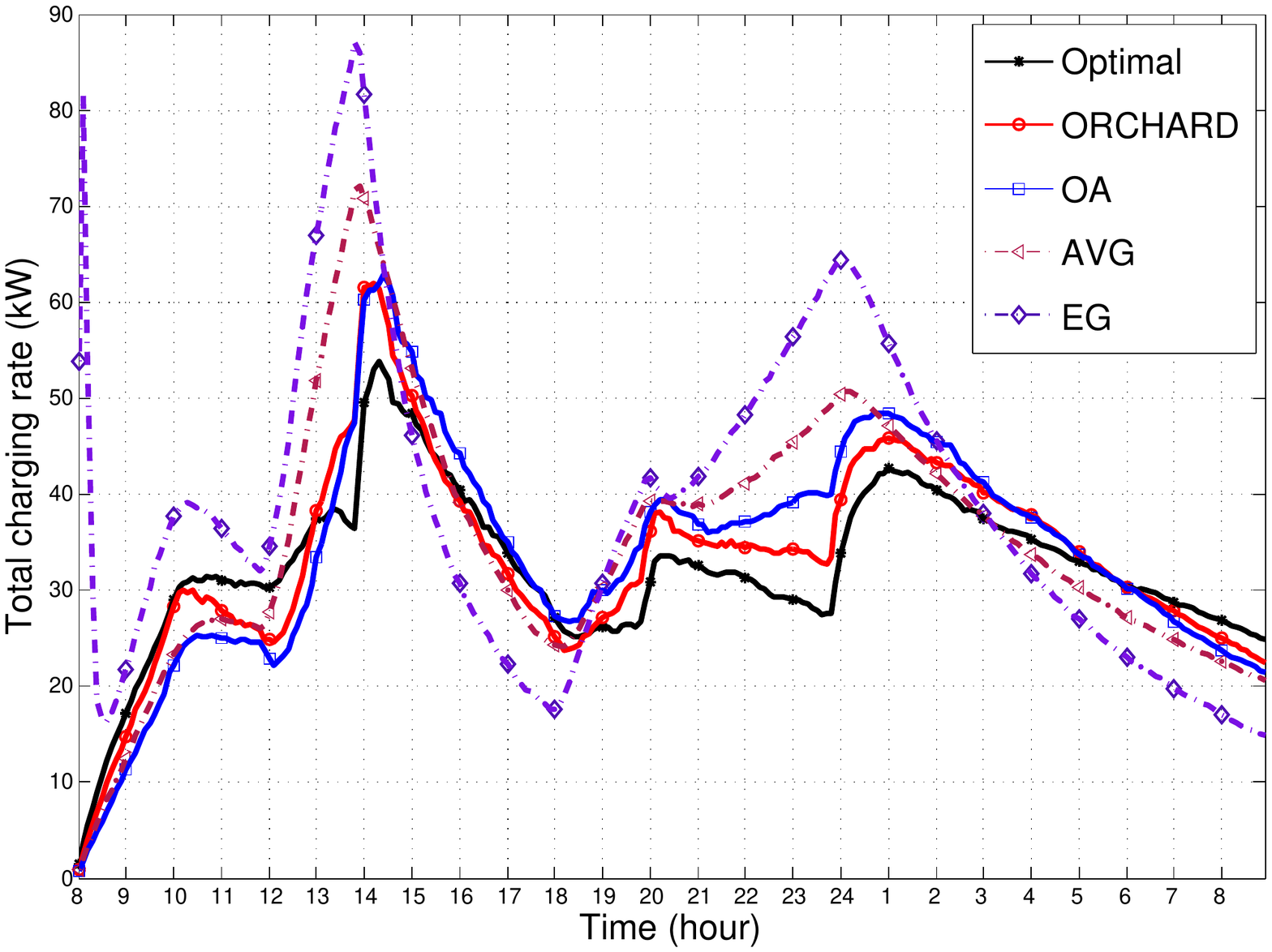}
        \label{fig:oneday_peak}
}
%\hspace{-0.05in}
\subfigure[Scenario 3: heavy traffic]{
    \includegraphics[width=0.31\textwidth]{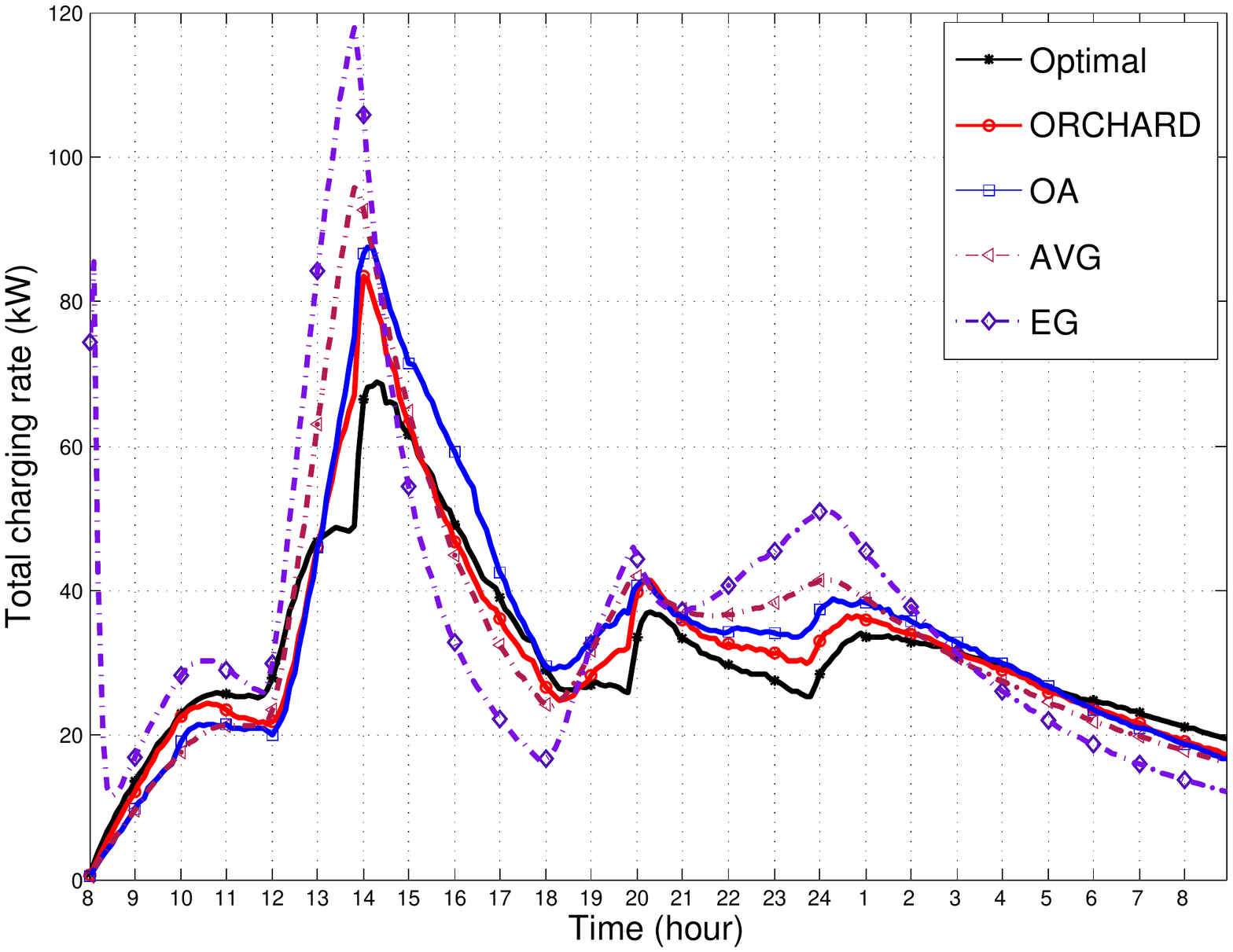}
        \label{fig:oneday_crowd}
}
\caption{
PEV total charging rate of five algorithms in three different scenarios.
%(a) Scenario two: total charging rate comparison of one day with light traffic
%(b) Scenario one: total charging rate comparison of one day with medium traffic
%(c) Scenario three: total charging rate comparison of one day with heavy traffic
\label{fig:charging_load}
}
%\vspace{-0.3cm}
\end{figure*}

\subsection{Performance Ratio Evaluation}\label{subsec:performance}
In this section, we evaluate the performance of ORCHARD.
We consider a running time $T$ of $24$ hours,
The coefficients of the cost function are set to
$a = 10^{-4} ~\text{\$/kWh}$ and $b = 0.6 \times 10^{-4} ~\text{\$/kWh/kW}$.
There are two types of PEVs in our simulation \cite{Ipakchi09grid}:
1)
maximum charging rate $U_i=3.3 kW$, battery capacity $\zeta_i=35 kWh$;
2)
maximum charging rate $U_i=1.4 kW$, battery capacity $\zeta_i=16 kWh$.
Each PEV is equally likely chosen from the two types
and the charging demand is uniformly chosen from $[0, \min \{ U_i\cdot (t_i^{(e)} - t_i^{(s)}), \zeta_i \}]$
(this ensures that (\ref{model6}) is feasible).
%We consider three different scenarios.
Each PEV's arrival follows a Poisson distribution and the parking time follows an Exponential distribution.
We consider three different scenarios, whose mean arrival and parking durations are listed in Table \ref{table:parameter}.
%The parameters of arrival rate and parking time are listed in Table \ref{table:parameter}.
In particular, Scenario one (S. 1), Scenario two (S. 2), and Scenario three (S. 3)
represent light traffic, moderate traffic and heavy traffic, respectively. The main difference lies in the arrival rates at the two peak hours, i.e. 12:00 to 14:00 and 18:00 to 20:00.
%We consider two scenarios in which one has obvious peaks in the arrival pattern while the other one has a relatively flat arrival pattern.

\begin{table}
\small		
        \caption{Parameter settings of the three scenarios}
        \centering
        %\begin{tabular}{r r r r r}
        \begin{tabular}{c c c c c}
        \toprule
         \multicolumn{1}{c}{\multirow {2}{*}{Time of Day}} & \multicolumn{3}{c}{Arrival Rate (PEVs/hour)} & \multicolumn{1}{c}{Mean Parking} \\
         \multicolumn{1}{c}{} & \multicolumn{1}{c}{S. 1} & \multicolumn{1}{c}{S. 2} & \multicolumn{1}{c}{S. 3} & \multicolumn{1}{c}{Time (hour)} \\
        \midrule
         $08:00 - 10:00$ &  $7$ & $7$ & $7$ & $10$ \\
         $10:00 - 12:00$ &  $5$ & $5$ & $5$ & $1/2$ \\
         $12:00 - 14:00$ &  $10$ & $30$ & $50$ & $2$ \\
         $14:00 - 18:00$ &  $5$ & $5$ & $5$ & $1/2$ \\
         $18:00 - 20:00$ &  $10$ & $30$ & $50$ & $2$ \\
         $20:00 - 24:00$ &  $5$ & $5$ & $5$ & $10$ \\
         $24:00 - 08:00$ &  $0$ & $0$ & $0$ & $0$ \\
         \bottomrule
        \end{tabular}
        \label{table:parameter}
        %\vspace{-0.4cm}
\end{table}

We compare ORCHARD to the optimal offline algorithm as well as other online algorithms.
Unless otherwise specified, the speeding factor of ORCHARD $q$ is set to be $1.46$.
We denote the cost of ORCHARD and the optimal offline algorithm by $\Psi_{ORC}$ and $\Psi^*$, respectively.
The other online algorithms for comparison are
\begin{enumerate}
  \item online average charging (AVG):
  	  The charging demand is evenly distributed during the parking period,
  	  i.e. the charging rate is $D_i/(t_i^{(e)} - t_i^{(s)})$.
  \item online eagerly charging (EG):
  	  PEV $i$ is charged at the maximum charging rate $U_i$.
  \item online optimal available information charging (OA)
  	  %(adapted from \cite{he2012optimal})
    : Set $q=1$ in ORCHARD.
\end{enumerate}
Their costs are denoted by $\Psi_{AVG}$, $\Psi_{EG}$ and $\Psi_{OA}$, respectively.
All the convex optimizations are solved by CVX\cite{grant2013}.

%\subsubsection{Performance Analysis}
For each scenario, we simulate $10^5$ cases and
plot the average total charging rate over time in Fig.~\ref{fig:charging_load}.
Besides, the average performance ratios normalized against the optimal offline solution are shown in Table \ref{table:ratio}.
In all scenarios, ORCHARD works the best among the four online algorithms, which has on average less than $14\%$ extra cost compared with the optimal offline algorithm.
We also notice that ORCHARD has a $10\%$ performance gain compared with the OA algorithm in the scenario with heavy traffic.
%which is achieved by setting $q=1.46$ instead of $q=1$.
We will discuss the proper setting of $q$ in Section \ref{subsec:discussion_q}.
The charging rate curve of the proposed online charging algorithm follows closely with the optimal offline solution curve.
In contrast, EG and AVG largely deviate from the optimal charging curve, being either too aggressive or too conservative depending on the arrival patterns.
In general, all charging algorithms perform better when the the traffic is relatively light, except for EG.
It produces even the worst performance ratio under light traffic. This is partly because its aggressive charging scheme somehow matches with the large traffic variations in scenario $3$.

\begin{table}
\small
        \caption{Average normalized performance ratio of online algorithms}
        \centering
        \begin{tabular}{c c c c c}
        \toprule
        Scenario & $\frac{ \Psi_{ORC} } { \Psi^* } $ & $ \frac{ \Psi_{OA} } { \Psi^* } $ & $ \frac{ \Psi_{AVG} } { \Psi^* } $ & $\frac{ \Psi_{EG} } { \Psi^* } $ \\
        \midrule
        1 & 1.068 & 1.135 & 1.530 & 2.346 \\
        2 & 1.104 & 1.197 & 1.645 & 2.309 \\
        3 & 1.133 & 1.240 & 1.701& 2.273 \\
         \bottomrule
        \end{tabular}
        \label{table:ratio}
\end{table}

\subsection{Setting a Proper $q$}\label{subsec:discussion_q}
\begin{figure}
        \centering
        \includegraphics[width=2.3in]{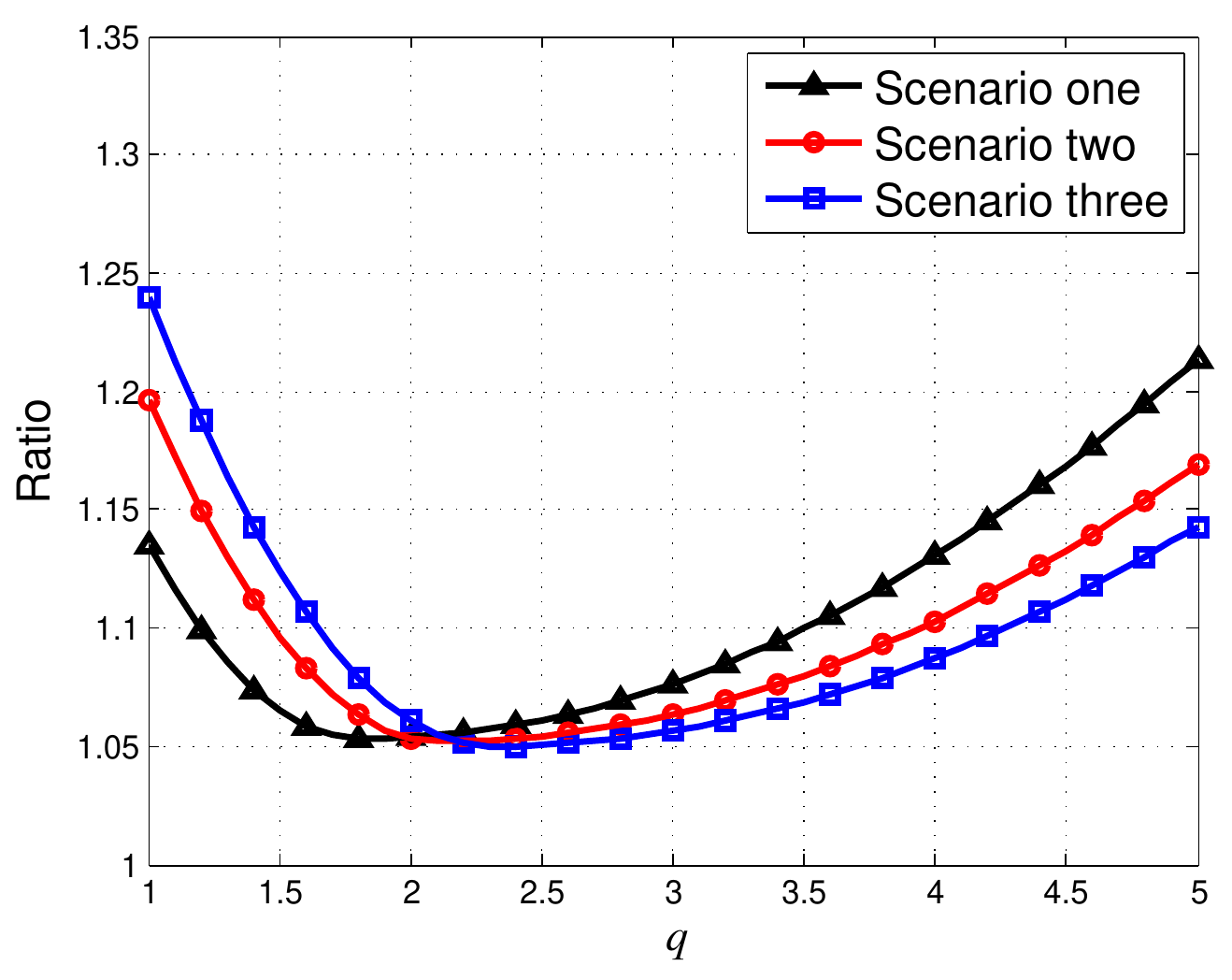}
        \caption{Average performance ratios of ORCHARD in two scenarios with varied $q$}
        %\vspace{-0.6cm}
        \label{fig:ratio_q}
\end{figure}

Theoretically, setting $q$ to be $1.46$ will achieve the best ratio in the worst case.
However, it does not achieve the best average performance in general.
In this subsection, we discuss how $q$ affects the normalized average performance ratio.
For the three scenarios with different traffic, we plot the normalized average performance ratio in Fig.~\ref{fig:ratio_q} by varying $q$ from $1$ to $5$.
For scenario $1$, setting $q=1.8$, $\frac{ \Psi_{ORC} } { \Psi^* }$ achieves the lowest average ratio $1.053$. For scenario $2$, setting $q=2.1$, $\frac{ \Psi_{ORC} } { \Psi^* }$ achieves the lowest average ratio $1.052$. For scenario $3$, setting $q=2.3$, $\frac{ \Psi_{ORC} } { \Psi^* }$ achieves the lowest average ratio $1.050$, which is about $8\%$ lower than that when $q=1.46$. In general, the optimal $q$ is larger when the traffic is heavy and unpredictable as in scenario $3$. Intuitively, this is because the charging cost during peak arrivals largely dominates the overall cost. A larger $q$ is able to better utilize off-peak hour and to speed up charging when peak hours arrive. From empirical results, if the peak load is about $2$ times of the flat load, $q$ is chosen to be $1.8$; if the peak load is about $6$ times of the flat load, $q$ is chosen to be $2.1$; if the peak load is about $10$ times of the flat load, $q$ is chosen to be $2.3$.
%\subsection{Complexity of optimal balanced charging algorithm}
%To verify the iteration complexity of the optimal balanced charging algorithm, we adopt it to solve program (\ref{model6}). For the system parameter, we reserve the same settings with the default settings except the arrival rates, which are assumed to be the same during $8:00 - 18:00$ and $0$ after $18:00$. We vary the arrival rate in $8:00 - 18:00$ from $1$ to $10$ (PEVs/hour) that leads to the mean of $N$(the number of total PEVs one day) varies from $10$ to $100$. For each specified mean of $N$, we simulate 10000 times and compute the average number of iterations and operations. The result is shown in Fig. \ref{fig:offlinecomplexity}. We also fit the data of iterations and operations with the polynomial function $f(x) = 2019.5x^4 - 297.9x^3 + 18.6x^2 - 0.2x$ and $f(x) = 166010x^5 - 26270x^4 + 1200x^3 - 10x^2$ with the \emph{mean relative error} $0.039$ and $0.054$ respectively. It shows that the complexity of both iteration and operation match our analysis.
%
%\begin{figure}
%        \centering
%        \includegraphics[width=2.3in]{OfflineComplexityEV.pdf}
%        \caption{Iteration number of optimal balanced charging algorithm}
%        \vspace{-0.5cm}
%        \label{fig:offlinecomplexity}
%\end{figure}

\section{Conclusions}\label{sec:conclusions}
In this paper, we have proposed an Online cooRdinated CHARging Decision (ORCHARD) algorithm,
which minimizes the energy cost without knowing the future information.
Through rigorous proof, we showed that ORCHARD is strictly feasible in the sense that it guarantees to fulfill all charging demands before due time. Meanwhile, it achieves the best known competitive ratio of 2.39.
To further reduce the computational complexity of the algorithm, we proposed a novel reduced-complexity algorithm to replace the standard convex optimization techniques used in ORCHARD.
Through extensive simulations,
we showed that the average performance gap between ORCHARD and the optimal offline solution, which utilizes the complete future information, is as small as $14\%$.
By setting proper speeding factor, the average performance gap can be further reduced to less than $6\%$.


\begin{thebibliography}{1}

\bibitem{sovacool2009beyond}
B.~K.~Sovacool, R.~F.~Hirsh, ``Beyond batteries: An examination of the benefits and barriers to plug-in hybrid electric vehicles (PHEVs) and a vehicle-to-grid (V2G) transition'', \emph{Energy Policy}, vol. 37, no. 3, pp. 1095-1103, 2009.

\bibitem{lopes2011integration}
J.~A.~P.~Lopes, F.~J.~Soares, and P.~M.~R.~Almeida, ``Integration of electric vehicles in the electric power system'', \emph{Proc. of the IEEE}, vol. 99, no. 1, pp. 168-183, Jan. 2011.

\bibitem{sortomme2011coordinated}
E.~Sortomme, M.~M.~Hindi, S.~D.~J.~MacPherson and S.~S.~Venkata, ``Coordinated charging of plug-in hybrid electric vehicles to minimize distribution system losses'', \emph{IEEE Trans. Smart Grid}, vol.2, no.1, pp. 198-205, 2011.

\bibitem{ma2010decentralized}
Z.~Ma, D.~Callaway and I.~Hiskens, ``Decentralized charging control
for large populations of plug-in electric vehicles: Application of the
Nash certainty equivalence principlee'', \emph{in Proc. IEEE Int. Conf. Control Appl.}, Sep. 2010, pp. 191-195.

\bibitem{allan98}
B.~Allan, E.~Ran, \emph{Online Computation and Competitive Analysis}, Cambridge, U.K.: Cambridge Univ. Press, 1998.

\bibitem{gerding2011online}
E.~Gerding, V.~Robu, S.~Stein, D.~Parkes, A.~Rogers, N.~Jennings, ``Online Mechanism Design for Electric Vehicle Charging'', \emph{in Proc. of 10th Int. Conf. on Autonomous Agents and Multiagent Systems (AAMAS 2011)}, May 2011, pp. 811-818.

\bibitem{masoum2012distribution}
M.~A.~S.~Masoum, P.~S.~Moses , S.~Hajforoosh  ``Distribution Transformer Stress in Smart Grid with Coordinated Charging of Plug-In Electric Vehicles'', \emph{IEEE Power Energy Syst. Innovative Smart Grid Tech. Conf.}, 2012, 1-8.

\bibitem{clement2010impact}
K.~Clement-Nyns , E.~Haesen and J.~Driesen, ``The impact of Charging Plug-in Hybrid Electric Vehicles on a Residential Distribution Grid'',  \emph{IEEE Trans. Power Syst.}, vol. 25, no. 1, pp. 371-380, 2010.


\bibitem{gan2012optimal}
L.~Gan, U.~Topcu, S.~H.~Low, ``Optimal Decentralized Protocol for Electric Vehicle Charging'', \emph{IEEE Trans. on Power System}, vol.28, iss. 2, pp. 940-951, 2012.

\bibitem{he2012optimal}
Y.~He, B.~Venkatesh, L.~Guan, ``Optimal Scheduling for Charging and Discharging of Electric Vehicles'', \emph{IEEE Trans. on Smart Grid}, vol.3, no.3, pp. 1095-1105, 2012.

\bibitem{chen2012iems}
S.~Chen, L.~Tong, ``iEMS for Large Scale Charging of Electric Vehicles Architecture and Optimal Online Scheduling'', \emph{in Proc. IEEE Int. Conf. Smart Grid Commun. (SmartGridComm)}, Nov. 2012, pp. 629-634.

\bibitem{yao1995a}
F.~Yao, A.~Demers, S.~Shenker, ``A Scheduling Model for Reduced CPU Energy'', \emph{in Proc. IEEE Symp. Foundations of Computer Science}, 1995, pp. 374-382.

\bibitem{bansal2007speed}
N.~Bansal, T.~Kimbrel, K.~Pruhs, ``Speed Scaling to Manage Energy and Temperature'', \emph{Journal of the ACM (JACM)}, vol. 54, no. 1, pp. 1-39, 2007.

%\bibitem{pruhs2007competitive}
%K. Pruhs, ``Competitive online scheduling for server systems'', \emph{SIGMETRICS Performance Evaluation Review}, 2007.
%
%\bibitem{bansal2008scheduling}
%N.~Bansal, H.~L.~Chan, T.~W.~Lam, L.~K.~Lee, ``Scheduling for Speed Bounded Processors'', \emph{Automata, Languages and Programming}, 2008.
%
\bibitem{bansal2009improved}
N.~Bansal, H.~L.~Chan, K.~Pruhs, D.~Katz, ``Improved Bounds for Speed Scaling in Devices Obeying the Cube-Root Rule'', \emph{Proc. 36th Int. Colloqium on Automata, Languages and Programming: Part I}, Jul. 2009, pp. 144-155.

\bibitem{bansal2009speed}
N.~Bansal, H.~L.~Chan, K.~Pruhs, ``Speed Scaling with an Arbitrary Power Function'', \emph{In Proc. of the 20th ACM-SIAM Symposium on Discrete Algorithm}, 2009, pp. 693-701.

\bibitem{lam2009speed}
T.~W.~Lam, L.~K.~Lee, Isaac K. K. To, and Prudence W. H. Wong, ``Speed Scaling Functions for Flow Time Scheduling based on Active Job Count'', \emph{Algorithms-ESA 2008}, 2008, pp. 647-659.

\bibitem{kothari03}
D.~P.~Kothari, I.~J.~Nagrath, \emph{Modern Power System Analysis}, 2003 :McGraw-Hill.

\bibitem{boyd04}
S.~Boyd, L.~Vandenberghe, \emph{Convex Optimization}, Cambridge, U.K.: Cambridge Univ. Press, 2004.

\bibitem{ye97}
Y.~Ye, \emph{Interior Point Algorithms: Theory and Analysis}, Wiley-Interscience Press, 1997.

\bibitem{grant2013}
M.~Grant and S.~Boyd, CVX: Matlab Software for Disciplined Convex Programming [Online]. Available: http://cvxr.com/cvx Mar. 2013, Version 2.0 (beta).

\bibitem{Ipakchi09grid}
A.~Ipakchi and F.~Albuyeh, ``Grid of the future'', \emph{IEEE Power and Energy Mag.}, vol. 7, no. 2, pp. 52-62, 2009.

%\bibitem{tang13report}
%W.~Tang, S.~Bi and Y.~J.~Zhang, ``Online Coordinated Charging Decision Algorithm for Electric Vehicles without Future Information'', technique report, 2013.
\end{thebibliography}
\end{document}